\documentclass[final,5p,times,twocolumn]{elsarticle}

\usepackage{amsmath,amssymb,amsfonts}
\usepackage{algorithmic}
\usepackage{graphicx}
\usepackage{textcomp}
\usepackage{xcolor}
\usepackage{subfigure, booktabs, multirow, makecell}
\usepackage[export]{adjustbox}
\usepackage{url,tikz}

\journal{Applied Energy}

\begin{document}

\begin{frontmatter}


 
\newcommand\copyrighttext{%
	\small \begin{center} \color{red} \textcopyright\,2022 Elsevier. Personal use of this material is permitted. Permission from Elsevier must be obtained for all other uses, in any current or future media, including reprinting/republishing this material for advertising or promotional purposes, creating new collective works, for resale or redistribution to servers or lists, or reuse of any copyrighted component of this work in other works.\end{center}}
\newcommand\copyrightnotice{%
	\begin{tikzpicture}
	\node 
	{\color{red}\fbox{\parbox{\dimexpr\textwidth-\fboxsep-\fboxrule\relax}{\copyrighttext}}};
	\end{tikzpicture}%
}

\title{\copyrightnotice\\ Experimental Investigation of Variational Mode Decomposition and Deep Learning\\for Short-Term Multi-horizon Residential Electric Load Forecasting}


\author[inst0]{Mohamed Aymane Ahajjam}
\affiliation[inst0]{organization={School of Electrical Engineering and Computer Science, University of North Dakota},
            addressline={Grand Forks 58202, ND}, 
            country={USA}}
\affiliation[inst1]{organization={College of Engineering \& Architecture - TICLab, International University of Rabat},
            addressline={Technopolis Rabat-Shore}, 
            country={Morocco}}

\affiliation[inst2]{organization={ENSIAS, Mohammed V University in Rabat},
            city={Rabat},
            country={Morocco}}
\affiliation[inst3]{organization={Czech Technical University in Prague} ,
            city={Prague},
            country={Czech Republic}} 
\affiliation[inst4]{organization={School of EEE, Faculty of Engineering, University of Leeds}, 
            city={Leeds},
            country={UK}}
            
\author[inst3]{Daniel Bonilla Licea}
\author[inst1,inst4]{Mounir Ghogho}
\author[inst1,inst2]{Abdellatif Kobbane}

\begin{abstract}
With the booming growth of advanced digital technologies, it has become possible for users as well as distributors of energy to obtain detailed and timely information about the electricity consumption of households. These technologies can also be used to forecast the household's electricity consumption (a.k.a. the load). In this paper, Variational Mode Decomposition and deep learning techniques are investigated as a way to improve the accuracy of the load forecasting problem. Although this problem has been studied in the literature, selecting an appropriate decomposition level and a deep learning technique providing better forecasting performance have garnered comparatively less attention. This study bridges this gap by studying the effect of six decomposition levels and five distinct deep learning networks. The raw load profiles are first decomposed into intrinsic mode functions using the Variational Mode Decomposition in order to mitigate their non-stationary aspect. Then, day, hour, and past electricity consumption data are fed as a three-dimensional input sequence to a four-level Wavelet Decomposition Network model. Finally, the forecast sequences related to the different intrinsic mode functions are combined to form the aggregate forecast sequence. The proposed method was assessed using load profiles of five Moroccan households from the Moroccan buildings' electricity consumption dataset (MORED) and was benchmarked against state-of-the-art time-series models and a baseline persistence model.
\end{abstract}

\begin{highlights}
    \item Propose a VMD-mWDN technique for short-term multi-horizon residential load forecasting.   
    \item Investigate the impact of decomposition levels on forecasting performance.
    \item Performance benchmarking using multiple deep learning architectures and under different scenarios.
    \item Technical experiments are carried out using load profiles acquired from five diverse Moroccan households.
\end{highlights}


\end{frontmatter}

\section{Introduction}

In smart grids, energy management strategies are aimed at monitoring, optimizing, and controlling energy consumption \cite{EnergyManadement}. Load forecasting plays an important role in this context as it allows utilities to make informed decisions about energy supply, and thus improve their energy generation and distribution. Indeed, more accurate load forecasts, resulting from advanced forecasting techniques, can significantly improve the efficiency of energy planning and supply operations.

Electricity consumption is a stochastic process with a degree of regularity that occurs due to human behavioral responses to such factors as day-night cycles and weekend patterns \cite{Predictibility2010}. The regularity of such patterns is critical to forecasting energy consumption.

This forecasting task can be studied in the context of residential, commercial, or industrial settings. The latter settings typically operate according to a predetermined schedule. While individual households can exhibit higher variability in their electricity consumption because residents' behavior is fairly improvised and not preplanned. Thus, forecasting the electricity consumption in residential settings is more challenging than in other settings. Nevertheless, the residents' consumption can be influenced by various known factors such as their demographic attributes (e.g., socioeconomic status, employment status), physical features of their premises (e.g., type of premises, size), and some exogenous parameters (e.g., weather and time of day) \cite{Transactions2020}.

As has been shown in the literature \cite{Kyriakides2007}, early works addressing load forecasting relied on time-series models \cite{TimeSeries2003}, regression-based techniques \cite{RegressionModels1997}, and Kalman filtering \cite{KalmanFilter1997}. However, the transition to advanced processing and machine learning techniques has proven to be very beneficial in handling complex load profiles and improving forecasting performances. Nevertheless, only few papers in the literature addressed short-term (i.e., hours to days) residential load forecasting, and even fewer papers dealt with multi-horizon residential load forecasting \cite{ReviewForecasts2020-2}. Yet, the recent availability of affordable and accurate smart meters, and their consequent adoption by residential premises, has enabled a much cheaper, easier, and convenient means of collecting consumption data, which can boost research on residential load forecasting as well as benefit both utilities and residents.        

\subsection{Literature review}
In the following, a literature review is conducted on residential load forecasting techniques focusing on three main factors: (i) forecasting in residential settings, (ii) multi-horizon load forecasting, (iii) and input decomposition techniques for load forecasting. 

\subsubsection{Load forecasting for residential settings}\label{Sec:1-2-1}
Residential load forecasting can be divided into two different problems. The first involves forecasting the electricity consumption of individual households independently of others. While the second aims at forecasting the aggregate consumption of a group of households relying on the smoothing effect resulting from this aggregation \cite{forecasting2016}. In this paper, the focus is on individual household forecasting. 

For instance, the authors of \cite{SingleHousehldForecasting2018} forecasted the electricity load of multiple Canadian households using one-hour granularity data acquired from a utility company. Exogenous features were generated to complement the load profiles, including weather conditions, temperature, humidity, season, day and time, and time of use prices. A Support Vector Regressor (SVR) was used to make single horizon forecasts using load consumption of hourly and daily granularities. Performance was evaluated using Mean Absolute Percentage Error (MAPE), where smaller values were achieved for the daily granularity. 

The authors of \cite{Lusis2017} studied the impact of calendar effects, forecasting granularity, and size of the training set in the task of single-horizon day-ahead forecasts of residential households. Several machine learning techniques were investigated: multiple linear regression, Regressions tree, SVR, and shallow neural networks. Performance was assessed using root mean square error (RMSE) and normalized RMSE, where regression trees using larger granularities provided the lowest values. The authors also concluded that calendar effects and large training sets of more than a year only slightly improve the forecasting performance. 

In \cite{Aurangzeb2019} the author compared the performance of eight regression models, including linear regression, kernel ridge regression, and SVRs with different kernels, for the task of short-term single-horizon forecasting of a single individual household. Performance was assessed using RMSE and MAE, with the smallest values achieved by SVR with a Radial basis function kernel. The data used in this study were obtained from the open dataset UCI Machine Learning Repository, reflecting the load consumption of a single household. Hence, the need for bigger datasets where the proposed models can be assessed on multiple households' consumption patterns.  

Recently, deep learning models have gained much attention for their capacity to learn complex features and to provide higher generalization capabilities than conventional machine learning algorithms. In \cite{Transactions2017}, the authors employed a deep Long Short-Term Memory (LSTM) network to forecast the single-horizon short-term electricity consumption of multiple Australian households from the open dataset SGSC. Three scenarios were considered reflecting one-hour, three-hours, and six-hours look-back time steps. Performance was assessed using MAPE and benchmarked against models including conventional back-propagation neural networks and k-nearest neighbor regression.

The authors of a recent study \cite{AppliedEnergy2021} developed a convolutional LSTM neural network model with selected autoregressive features for single-horizon short-term load forecasting over three different spatial granularities: apartment-, flour-, and whole building levels. This model's performance was benchmarked against various others, including SVRs, convolutional LSTM, persistence model, Auto-Regressive Integrated Moving Average model, using various inputs including temperature, absolute humidity, wind speed, and time. The forecasting accuracy was assessed based on the coefficient of variance (CV) metric.

In \cite{AppliedEnergy2021-2}, the authors proposed an online learning approach based on recurrent neural networks (modified LSTM network) for short-term single horizon load forecasting. Performance was assessed using MSE and MAE. Results showed the capability of the proposed model to learn in an online manner and achieve better forecasting results when compared with standard LSTM networks and five conventional machine learning techniques (e.g., linear regression, K-nearest neighbors). 

\subsubsection{Multi-horizon load forecasting}
Electric load forecasting can be carried out for single or multiple time horizons. In the former, a forecast is made for a single instant, while in the latter, a forecast is made for many time instants. Multi-horizon load forecasting enables the forecasting model to learn the temporal correlations between consecutive horizons. In the literature, few papers consider multi-horizon load forecasting. 

In \cite{Algeria2019}, the authors propose a two-stage forecasting model based on clustering similar daily load profiles followed by forecasting using multiple denoising autoencoders. Daily temperature data are estimated and incorporated in the proposed technique as a way to enhance its forecasting performance when actual temperature profiles are not available. Using lagged load values and temperature estimations, single-horizon hourly forecasts are employed in a recursive manner to generate full day forecasts. Performance was assessed using MAPE and MAE. This study relied on electricity consumption data collected over four years by Algeria’s National Electricity and Gas Company. 

In \cite{multihorizon2019}, the authors aimed at increasing the residential load forecasting performance by excluding the portion of load consumption coming from highly consuming appliances that can be attributed to the outside temperature of the household (i.e., heating and air-conditioning), and rather focusing on the residual load. Consequently, they define operation schedules before injecting the results to linear regression models for day-ahead forecasting with a 1h granularity. Performance was assessed using MAE, normalized MAE, and the correlation coefficient $R^2$ on a single Canadian household.

The authors of \cite{Transactions2018} proposed a two-stage ensemble strategy based on deep residual network (ResNet) models to perform multi-horizon day-ahead forecasts with a one-hour granularity. In addition, probabilistic forecasting can be achieved using the proposed model with Monte Carlo dropout. The proposed model relies on load and temperature data from two open datasets reflecting the load demand from two U.S. utility companies. Performance was compared with existing models in the literature using MAPE, where higher accuracy and generalization capability were achieved using the proposed model. 

A very recent paper is \cite{AppliedEnergy2021-3}, where the authors adapted the logistic mixture autoregressive model for vector inputs in order to enable multi-horizon day ahead forecasting. Such a model combines pattern clustering and forecasting using the expectation-maximization algorithm. A curve registration was integrated into the proposed model to remove the high variability in daily load profiles and produce better clustering performance. Additional machine learning models were benchmarked, including multi-layer perceptron, sequence to sequence LSTM, and a persistence model. All models were developed to make multi-step day-ahead forecasts using load data from two commercial buildings (libraries) from the U.S. (96 step forecasts) and the Republic of Korea (288 step forecasts). Performance was assessed using RMSE, MAPE, CV, and the forecast skill, with the proposed model achieving higher accuracy than all considered models.  

\subsubsection{Decomposition techniques for load forecasting}\label{Sec:1-2-2}

With the stochastic nature of residential load consumption, decomposition techniques are used in the literature for time-frequency analysis, extraction of intrinsic information, and removal of noise and redundant information. For instance, the authors of \cite{EMDLSTM2020} proposed a short-term forecasting technique based on the Empirical Mode Decomposition (EMD) technique to decompose the highly volatile and non-stationary load into 14 stationary Intrinsic Mode Functions (IMF), and a residue fed to a deep LSTM network. Performance was assessed using MAPE and RMSE, and multiple models were compared against the proposed network. Results showed that models trained on EMD-processed inputs provided higher performance than models that did not. In \cite{EMD2017}, the authors proposed an ensemble forecasting technique combining EMD and a deep belief network formed from two restricted Boltzmann machines and one shallow artificial neural network. In this work, EMD decomposes the input time-series into eight IMFs and a residue. Two forecasting horizons were studied, namely, half an hour (i.e., very short term) and 24h (i.e., short-term). Performance was assessed with load demand datasets from the Australian Energy Market Operator, using RMSE and MAPE. In addition, a performance benchmark was conducted using four machine learning models (e.g., support vector machine, random forest), and a persistence model, with and without EMD-decomposition. Results showed that the proposed EMD-based technique outperforms other models for both forecasting horizons, as well as highlighted the usefulness of deep learning models when dealing with nonlinear features and bigger forecasting horizons.

The authors of \cite{VMD2021} proposed incorporating VMD, a chaotic mapping mechanism, and the grey wolf optimizer algorithm within a support vector regression forecasting model. VMD was employed to decompose the input time-series into six high-frequency IMFs of reduced non-linearity and non-stationarity. Results showcased the effectiveness of VMD within the proposed technique in handling the single-horizon hourly forecasting problem. In \cite{VMDEMD2018}, the authors compared the performance of Variational Mode Decomposition (VMD) and EMD within an LSTM network forecasting technique for different horizons. With a decomposition level equal to 10 IMFs (and a residue), the proposed VMD-LSTM technique provides higher forecasting accuracy. Similarly, the authors of \cite{VMDLSTM2019} developed a VMD-LSTM technique but also employed a Bayesian Optimization Algorithm for short-term single-horizon load forecasting. The optimization technique was used for multiple tasks, including optimizing the VMD decomposition level and the extension of the data in the case of a correlation between the input load data and specific related data (i.e., temperature, dew point, humidity, and day type). All data used in this paper were received from a Chinese grid utility. 
Significant improvement was achieved with the proposed technique when compared with other conventional machine learning models (i.e, SVR, linear regression) and decomposition techniques (i.e., EMD, ensemble EMD). 

The authors of \cite{Wavelet2010} investigated wavelet decomposition and neural networks to achieve 24h load forecasting. More specifically, the proposed technique relied on correlation analysis to select similar day's load and the fourth-order Daubechies (db4) wavelet to extract low and high frequency components. Additional features such as weekday index and weather-related indexes (e.g., temperature, wind speed, cloud cover) are injected into two individual shallow neural networks to forecast a single component of next-day's load (i.e., high-frequency or low frequency components). In \cite{WT2020}, the authors proposed a probabilistic forecasting technique based on the relevance vector machine model and a wavelet transform for one-hour and 24h load demand forecasts. The wavelet transform is employed to filter out high-frequency noise, in order to smooth the abrupt changes in the time-series and improve the forecasting performance. The proposed technique outperformed classic time-series forecasting techniques (e.g.,  seasonal autoregressive integrated moving average) as well as machine learning-based technique (support vector machine and shallow artificial neural network). In \cite{AppliedEnergy2019}, the authors employed wavelet decomposition within their short-term load forecasting technique and assessed its performance using individual appliance and whole-house consumption data of a single household from the open dataset AMPds. A feature transformation process based on one-level wavelet decomposition and a collaborative representation transform was used to transform input load curves to a new feature space of useful and less redundant information. The authors relied on a recursive single-horizon forecasting technique to make next-24h forecasts. The proposed model is an LSTM network containing a single LSTM layer and a fully connected layer. Although individual appliance consumption requires an intrusive and costly acquisition process when compared with that required for whole-house consumption, results showed significant improvements in the forecasting accuracy when using both types of load consumption data, as well as when using the proposed feature extraction phase. 

\subsection{Motivation and contribution}
In most of the above VMD-based studies, the decomposition level is chosen arbitrarily, and its impact on performance is never evaluated. In addition, VMD only decomposes the input sequence within the Fourier spectrum, identifying frequencies without their associated temporal dimension. To improve forecasting performance, an additional method is required in order to extract temporal information related to the consumption patterns inherent in load profiles.

On the other hand, only a few studies have implemented multi-horizon forecasting strategies, as most rely on recursive single-horizon forecasting to gain multiple predictions. In addition, exogenous data describing socioeconomic or weather features are commonly used to improve forecasting performance. However, such data are not always available or require costly equipment to collect. 

Further, in the majority of the studies, the proposed techniques were evaluated using load data from a single household, multiple commercial buildings, or only simulations; effectively failing to assess the impact of different electricity consumption behaviors from diverse households on performance. 

This paper aims at addressing those issues, and its main contributions are as follows:
\begin{itemize}
    \item A technique based on VMD and wavelet-based Convolutional Neural Network (CNN) is proposed for short-term multi-horizon load forecasting. The use of both decomposition techniques allows the extraction of spectral and temporal information reflecting various consumption behaviors from load profiles. This allows improved forecasting performance especially when exogenous data are not available.
    \item A study on the effects of decomposition levels on decomposition-based load forecasting performance is conducted to identify the optimal decomposition level achieving best results.
    \item Technical experiments are carried out using whole-house electricity consumption of five Moroccan households acquired during a data acquisition campaign set up during 2020 and 2021. Thus, providing different lifestyles and consumption patterns on which to evaluate the proposed technique. 
\end{itemize}

\subsection{Organization of the paper}

The rest of the paper is organized as follows. Section \ref{Sec:2} provides a brief background on the proposed deep learning-based forecasting technique, including VMD and Multilevel Wavelet Decomposition Network (mWDN). Section \ref{Sec:3} describes the technical aspect of the proposed multi-horizon forecasting process, the data acquisition campaign, and the benchmarking setup. Results of various experiments are discussed in Section \ref{Sec:4}. While Section \ref{Sec:5} summarizes and concludes the paper.



\section{Hybrid multi-horizon load forecasting using VMD and mWDN}\label{Sec:2}
There are various strategies to forecasting multiple horizons in the literature. These strategies can be categorized into direct and indirect approaches. Indirect approaches are characterized by relying on multiple single-horizon forecasts to produce multi-horizon forecasting.  

\begin{figure}[t]
    \centering
    \subfigure[Recursive single-horizon forecasting]
    {
        \includegraphics[width=0.85\columnwidth]{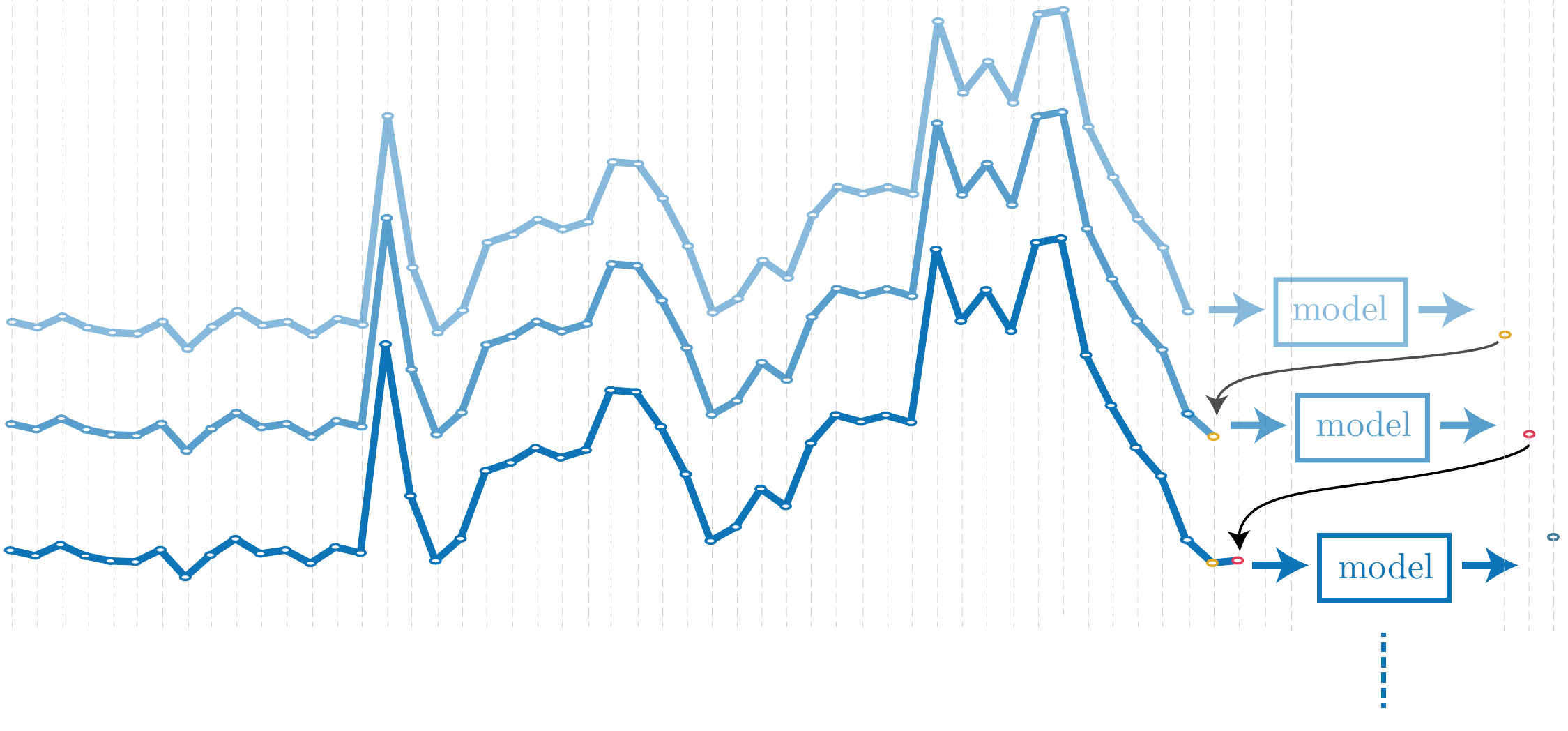}
        \label{fig:strategy1}
    }
    \\
    \subfigure[Direct single-horizon forecasting]
    {
        \includegraphics[width=0.85\columnwidth]{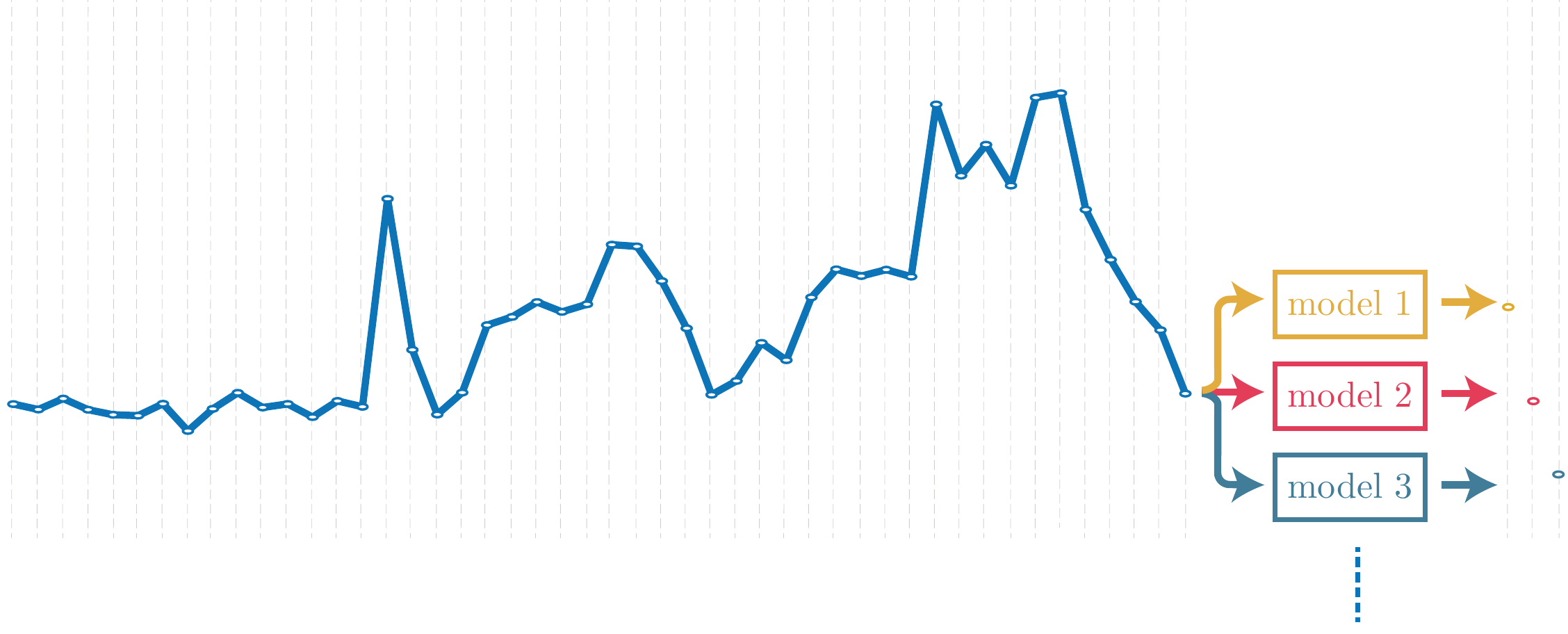}
        \label{fig:strategy2}
    }
    \caption{Two strategies for multi-horizon forecasting.}
    \label{fig:ForecastingStrategies}
\end{figure}

One indirect strategy to achieve multi-horizon forecasting is recursive single-horizon forecasting. This strategy relies on a single forecasting model that is trained to predict the next step, which in turn is fed back to it to predict the following step. This loop is repeated in order to have predictions over the desired number of horizons (see Fig.\ref{fig:strategy1}). However, this strategy suffers from a propagating error that accumulates over time, despite being simple and computationally inexpensive. 

Another strategy to produce multi-horizon forecasts is direct single-horizon forecasting. This strategy relies on multiple forecasting models, where each one is assigned to forecast a single-horizon (see Fig.\ref{fig:strategy2}). Although no error propagation occurs in this strategy, each model is trained for a single forecasting horizon. Therefore, it effectively ignores the correlation between successive forecasting horizons and requires computation costs that proportionally increase with the number of forecasting horizons.

Direct multi-horizon forecasting addresses the issues described above. More specifically, this strategy relies on a single forecasting model, as it is trained to forecast multiple steps at once. Nevertheless, this strategy is characterized by an increasing level of forecasting uncertainty with every horizon \cite{Taieb2016}.  In this article, the proposed forecasting technique implements this forecasting approach to generate multi-horizon forecasts of residential electricity consumption.  More specifically, the proposed forecasting technique considers the relationships between successive horizons to directly output 48 forecasts reflecting the next 24h load consumption. 

\begin{figure}[t]
  \centering
    \includegraphics[width=.98\columnwidth]{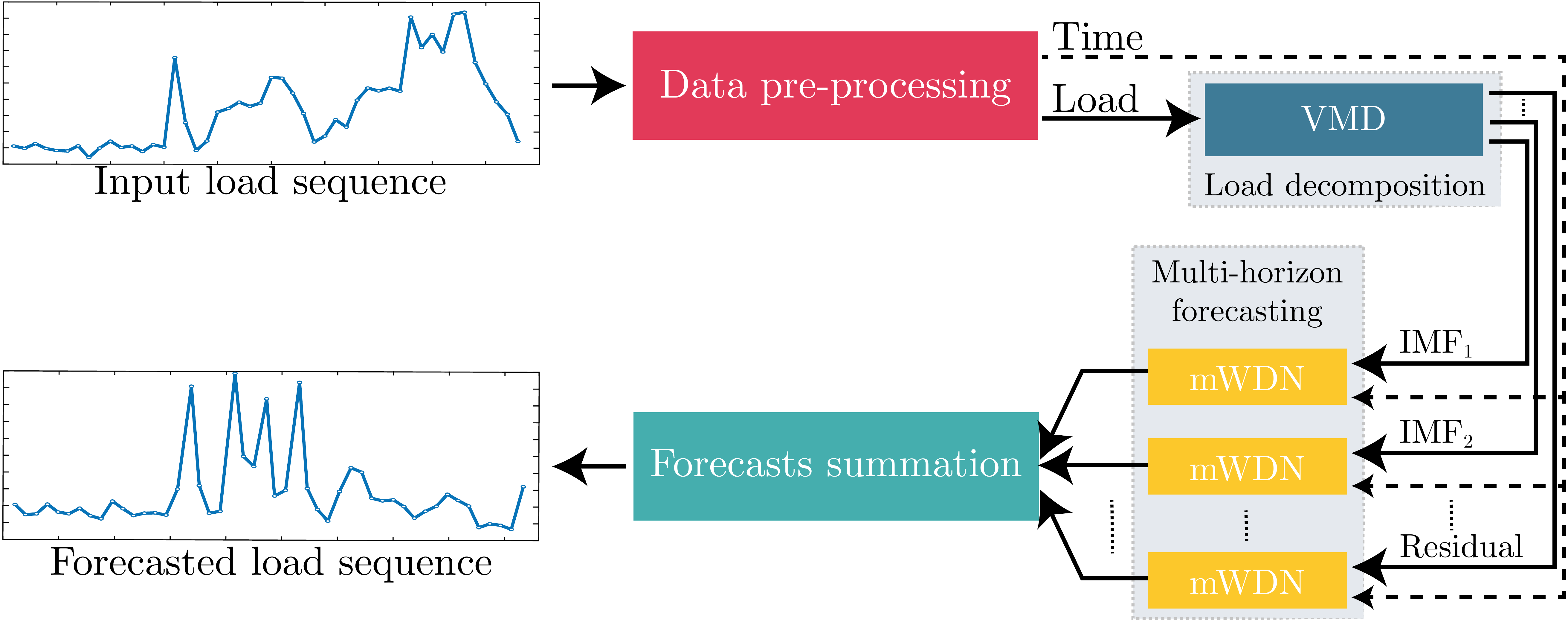}
    \caption{The proposed forecasting process.}\label{fig:Process}
\end{figure}

However, the proposed technique follows a "divide and conquer" approach (see Fig. \ref{fig:Process}), following four main steps \cite{VMDEMD2018}: 
\begin{itemize}
    \item Data pre-processing, where raw load data are averaged, reshaped, and normalized to produce sequences reflecting load and time information over 24h periods. 
    \item Load decomposition, where a pre-processed load sequence is decomposed using VMD to $K$ stationary sequences and a residue sequence.
    \item Multi-horizon forecasting, where $K+1$ wavelet-based deep learning models (i.e., mWDNs) take the $K+1$ decomposed load sequences (including the residue) and their corresponding time sequences to provide their corresponding forecasts.
    \item Forecasts summation, where the $K+1$ forecasted sequences of all $K+1$ models are summed to provide the actual forecast of the original input load sequence. 
\end{itemize}

In other words, the proposed forecasting technique takes as input a load sequence and generates its next 24h forecast over 48 horizons of 30 minutes duration each. However, the input load sequence is decomposed using VMD into $K+1$ decomposed load sequences under specific characteristics. The proposed forecasting technique uses $K+1$ wavelet-based deep learning models to cover all decompositions. Each model takes as input one decomposed load sequence (along with its corresponding time sequences), and is assigned to forecast the next 24h state of that decomposed load sequence (see Fig. \ref{fig:IMF2forecast}). These models are able to extract multi-spectral features using the wavelet decomposition and incorporate them into their training process to improve their forecasting performance. The summation of all forecasts form the actual forecast of the original input load sequence. 

A brief overview of the decomposition technique and forecasting model follows.

\subsection{Background and motivation of VMD}
\begin{figure*}[!h]
  \centering
    \includegraphics[width=.85\textwidth]{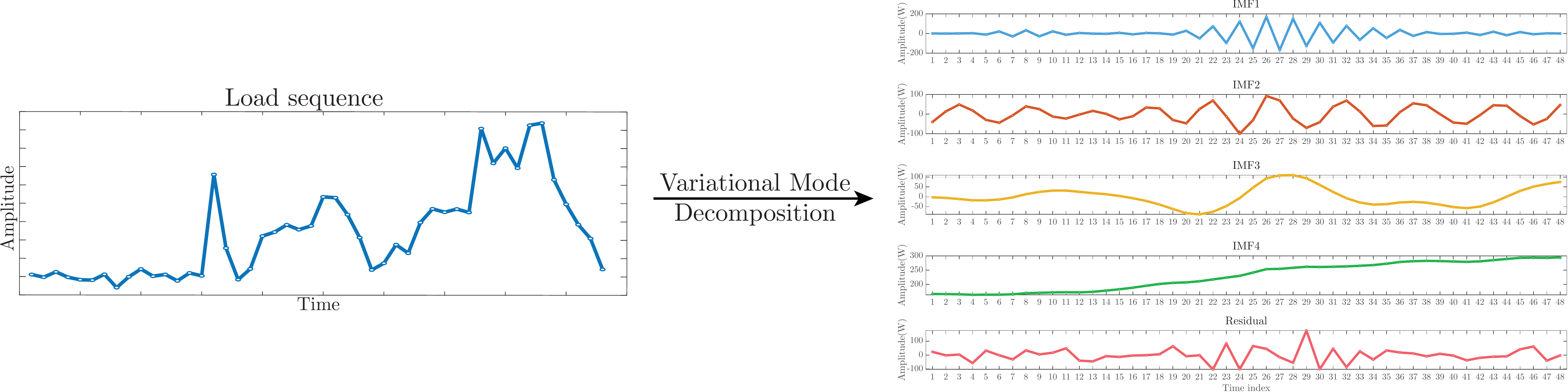}
    \caption{Decomposition of a load profile using VMD (K=4).}\label{fig:signal2IMFs}
\end{figure*}

In many fields, decomposition methods are widely used for the purpose of performing time-series processing such as noise removal \cite{NoiseRemovalEMD2020}, demodulation analysis \cite{DemodulationAnalysis2016}, and the extraction of inherent information \cite{BearingVMD2020}.
VMD differs from prior methods (i.e., Empirical Mode Decomposition) in that it is a mathematically framed decomposition technique \cite{VMD2013}. It is able to simultaneously decompose an input signal into a discrete number of inherently narrow-band and stationary signals (i.e., IMFs), while retaining the possibility of its reconstruction (see Fig.\ref{fig:signal2IMFs}). 

Let $\{E(t)\}_{t=1}^{H}$ denote a typically non-stationary electricity consumption sequence of H discrete values sampled at periodic intervals. Then, its decomposition using VMD can be written as:
\begin{equation}
 E(t) = \sum_{k=1}^{K}F_k(t) + r(t)
 \label{eq:IMFdecomposition}
\end{equation}
where $F_k(t)$ is the $k$th IMF sequence, $K$ is the decomposition level, and $r(t)$ is the residual. 

Based on \cite{VMD2013}, an IMF is an amplitude-modulated and frequency-modulated signal following:
\begin{equation}
 F_k(t) = A_k(t)\cos{\phi_k(t)}, \quad A_k(t)\ge 0
 \label{eq:IMF_expr}
\end{equation}
where $\phi_k(t)$ is the phase and $A_k(t)$ is the slowly changing envelope corresponding to the $k$th $IMF$; It also has a non-decreasing instantaneous frequency that varies slowly and is mostly compact around a center frequency $\omega_k$. 

The VMD algorithm performs several computations to find the K IMFs and their corresponding central frequencies concurrently via the optimization technique: Alternate Direction Method of Multipliers (ADMM)\cite{ADMM2011}. According to ADMM, VMD can decompose the input $E(t)$ into $K$ $F_k$ and $\omega_k$ using these equations\cite{VMD2013}:
\begin{equation}
 \mathcal{F}\big\{F_k^{n+1}\big\} = \frac{\mathcal{F}\{E(\omega)\}-\sum_{i\neq k}\mathcal{F}\big\{F^{n+1}_{i}(\omega)\big\}+\frac{\mathcal{F}\{\lambda^n(\omega)\}}{2}}{1+2\alpha(\omega-\omega_k^n)^2}
 \label{eq:IMF_update}
\end{equation}

\begin{equation}
 \omega_k^{n+1} = \frac{\int_0^\infty{\omega\big\lvert\mathcal{F}\big\{F_k^{n+1}(\omega)\big\}\big\rvert^2 d\omega}}{\int_0^\infty{\big\lvert\mathcal{F}\big\{F_k^{n+1}(\omega)\big\}\big\rvert^2 d\omega}}
 \label{eq:centerfrequency_update}
\end{equation}
where $n$ is the number of iterations, $\lambda$ is the Lagrangian multiplier, $\mathcal{F}\big\{F_k^{n+1}\big\}$, $\mathcal{F}\{E(\omega)\}$, $\mathcal{F}\{F(\omega)\}$, and $\mathcal{F}\{\lambda^n(\omega)\}$ correspond to the Fourrier transform of $F_k^{n+1}$, $E(t)$, $F(t)$, and $\lambda^n$, respectively. The initial value of $n$, as well as of other parameters, $\lambda^1$, $\mathcal{F}\big\{F_k^{1}\big\}$,
and $\omega_k^1$, are set to 0.

The decomposition level $K$ is considered a key parameter of VMD as a too small or a too big value can produce erroneous IMFs (e.g., overlapping, noisy). A suitable value must therefore be identified within the context of the task under consideration.  

\subsection{Background and motivation of mWDN}

\begin{figure*}[!t]
  \centering
    \includegraphics[width=.75\textwidth]{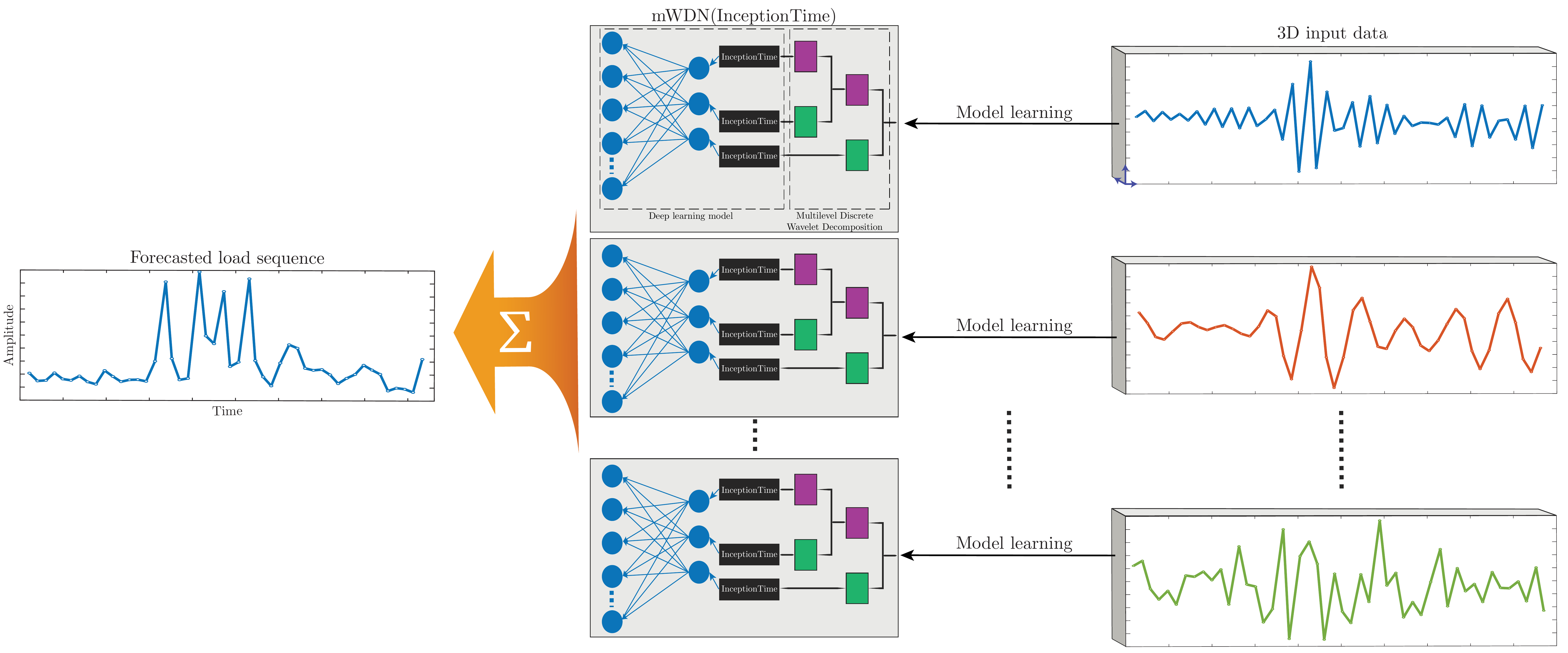}
    \caption{Load forecasting based on mWDN relying on InceptionTime models and a wavelet decomposition level of $I=2$.}\label{fig:IMF2forecast}
\end{figure*}

Wavelet decomposition techniques have proved beneficial in time-frequency analysis and provide a better alternative to Fourier transforms, as they can simultaneously extract local spectral and temporal information by relying on a window with variable widths\cite{WaveletReview}. Thus, enabling a local scale-dependent analysis of intrinsic consumption behaviors.

mWDN is a wavelet-based neural network structure that was introduced in 2018 in order to construct frequency-aware deep learning models\cite{wang2018}. This model is two parts: a time-frequency decomposition (i.e., wavelet decomposition) and a deep learning model (see Fig. \ref{fig:IMF2forecast}). Hence, mWDN can seamlessly integrate the Multilevel Discrete Wavelet Transform in a deep learning framework ensuring the fine-tuning of all parameters in the training phase. 

This model implements the standard wavelet decomposition approach where a signal is decomposed to high- and low-frequency sub-series (i.e., detail and approximation coefficients), which are further decomposed using latter sub-series following a number of decomposition levels. 

Let $\{e_n^l(i)\}_{n=1}^{N}$ and $\{e_n^h(i)\}_{n=1}^{N}$ denote the low and high sequences extracted from the input signal $\{e(t)\}_{t=1}^{T}$ in the $i$th decomposition level of a multilevel wavelet decomposition. Each sequence is generated using low and high pass filters $l=\{l_1,..,l_K\}$, $h=\{h_1,..,h_K\}$ followed by a downsampling technique (i.e., average pooling). Convolving $e^l(i)$ with $l$ and $h$ generates intermediate sequences $\{a_n^l(i+1)\}_{n=1}^{N/2}$ and $\{a_n^h(i+1)\}_{n=1}^{N/2}$ that can be expressed as:
\begin{equation}
 a_n^l(i+1) = \sum_{k=1}^{K}e_{n+k-1}^l(i).l_k
 \label{eq:lowpass_result}
\end{equation}
\begin{equation}
 a_n^h(i+1) = \sum_{k=1}^{K}e_{n+k-1}^l(i).h_k
 \label{eq:highpass_result}
\end{equation}
where $e_{n}^l(i)$ is the $n$th element of $e_{n}^l(i)$, with $e_{n}^l(0)$ corresponding to the input of the model. The term $N/2$ refers to the 1/2 downsampling of the intermediate sequences. 

Similarly, mWDN decomposes the input sequence using : 
\begin{equation}
 a^l(i+1) = \sigma\bigg(W^l(i+1)e^l(i)+b^l(i+1)\bigg)
 \label{eq:lowpass_mWDN}
\end{equation}
\begin{equation}
a^h(i+1) = \sigma\bigg(W^h(i+1)e^l(i)+b^h(i+1)\bigg)
\label{eq:highpass_mWDN}
\end{equation}
where $\sigma(.)$ is a sigmoid function, $W^l$ and $W^h$ are weight matrices, $e^l(i)$ and $e^h(i)$ the extracted low and high frequency sequences in the $i$th level. An average pooling layer of stride 2 with kernel size 2 is used to downsample from $a^h(i+1)$ and $a^l(i+1)$. 

Following Fig.\ref{fig:IMF2forecast}, the decomposition results after $I$ levels \Big(i.e., $\Big\{e^h(1),e^h(2),...,e^h(I),e^l(I)\Big\}$\Big) are injected in I+1 independent models, where each model is trained to forecast the future state of its input sequence. The actual forecast of the whole mWDN model is constructed using a fully connected neural network. 

Consequently, the proposed technique can effectively incorporate the benefits of both VMD and wavelet decomposition within a deep learning framework to compensate to a certain extent for the lack of exogenous features and improve the forecasting performance. Further details about the implementation of the proposed technique and the learning process are presented in the next section.

\section{Technical Implementation}\label{Sec:3}
\begin{figure}[!h]
  \centering
    \includegraphics[width=.45\textwidth]{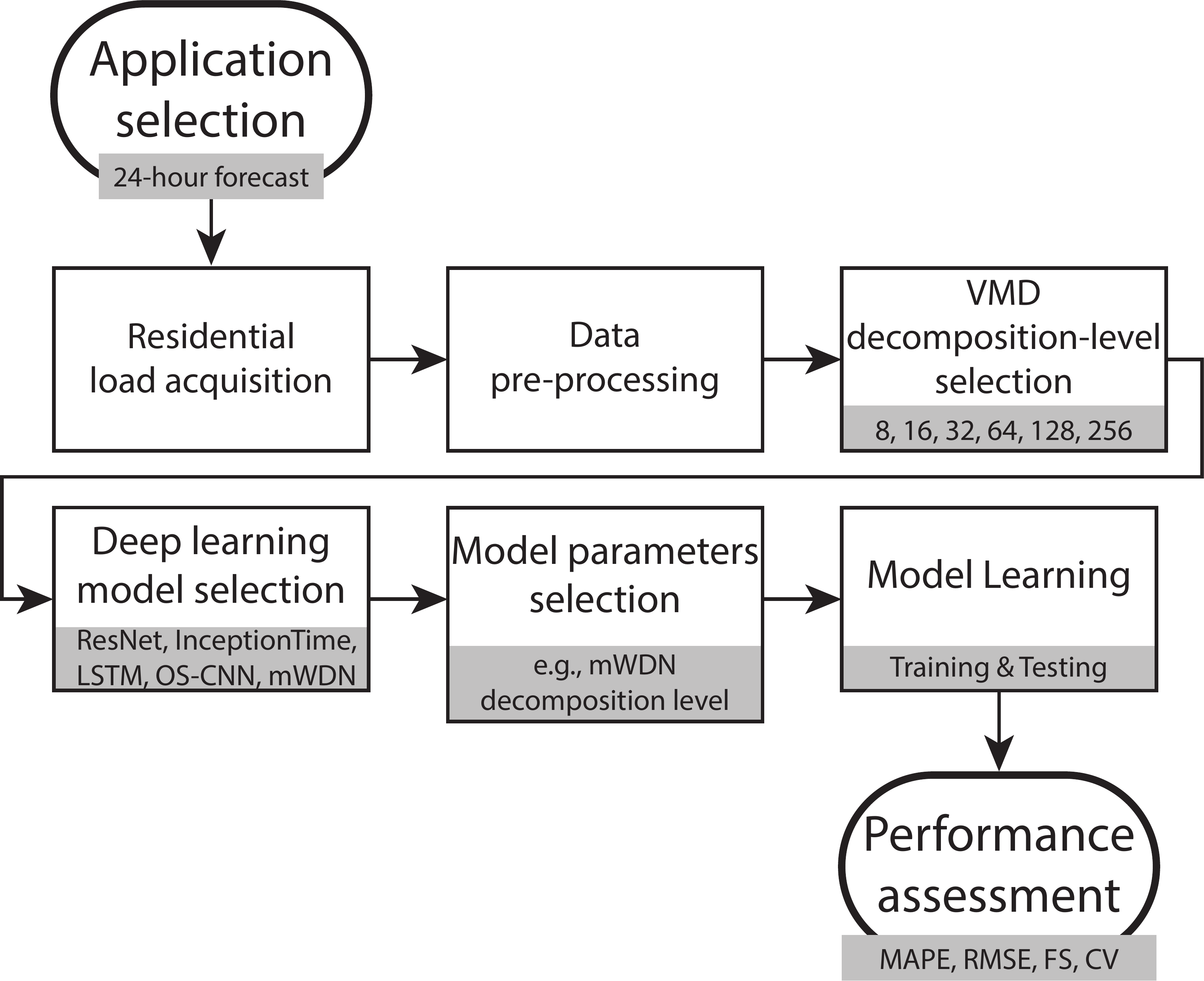}
    \caption{Forecasting model development.}\label{fig:ForecastModelSteps}
\end{figure}

The purpose of this section is to present the technical aspects of this study, starting with a description of VMD-based forecasting and its input data: electric load profiles acquired from Moroccan households. The last subsection details the benchmarking setup carried out to assess the performance of the proposed technique against state-of-the-art models. Fig.\ref{fig:ForecastModelSteps} highlights the major steps considered in this study to develop forecasting models. 

\subsection{VMD-empowered electricity consumption forecasting}\label{Subsection:VMDemporedForecasting}
Multi-horizon forecasting of an electricity consumption sequence $E_{T-S+1|T}=\big\{E(T-S+1),..,E(T)\big\}$, of $S-1$ elements and a time index $T$, can be written as:
\begin{equation}
 \hat{E}_{T+S|T+1} = f\big(E_{T-S+1|T}, \epsilon\big)
 \label{eq:forecasting}
\end{equation}
where $\hat{E}_{T+S|T+1}$ is the forecasted load consumption sequence over $S$-steps (i.e., horizons), $f$ is the forecasting model, $S$ is the horizon, and $\epsilon$ is the error.

The primary motivation of this work is to forecast the $\hat{E}_{T+S|T+1}$ over short horizons $S$ using load-related variables only.
For this end, the raw electricity consumption data are processed to multi-variate continuous 48h sequences, where the first 24h are fed to the model to forecast the last 24h (see Fig.\ref{fig:Process}).
Two main types of sequences are extracted:
\begin{itemize}
    \item Load sequence $E_{T-S+1|T}$: to provide direct historical 24h electricity consumption data with a granularity of 30 minutes (i.e., a sequence of 48 elements).   
    \item Time sequences : to provide recurrent temporal information comprising the corresponding values of hour of day $H_{T-S+1|T}=\big\{H(T-S+1),..,H(T)\big\}$ ranging from 0 to 23 and day of week $D_{T-S+1|T}=\big\{D(T-S+1),..,D(T)\big\}$ ranging from 0 to 6, with the same granularity (i.e., two sequences of 48 elements each).   
\end{itemize}

\begin{table*}[!t]
\centering
\caption{Hyper-parameters of the proposed model.}
\label{tab:ParametersModels}
\resizebox{\textwidth}{!}{%
\begin{tabular}{@{}llllll@{}}
\toprule
Model               & Configuration            &                      &                    &                     &                          \\ \midrule
mWDN(InceptionTime) & \textbf{mWDN}            &                      &                    &                     &                          \\
                    & Level = 4                &                      &                    &                     &                          \\
                    & wavelet = db4            & \multicolumn{1}{r}{} &                    &                     &                          \\
                    & Pooling layer            & Average pooling      &                    &                     &                          \\
                    &                          & kernel size = 3      &                    &                     &                          \\
                    &                          & stride = 1           &                    &                     &                          \\
                    & \textbf{InceptionTime}   & Inception modules    & 6                  &                     &                          \\
                    &                          &                      & Convolution layers & 3                   &                          \\
                    &                          &                      &                    & \multicolumn{2}{l}{filters = 32,32,32}         \\
                    &                          &                      &                    & \multicolumn{2}{l}{kernel sizes = 39,19,9}     \\
                    &                          &                      &                    & Pooling layer       & Max-pooling              \\
                    &                          &                      &                    &                     & kernel size = 3          \\
                    &                          &                      &                    &                     & stride = 1               \\
                    &                          &                      &                    & Convolution layer   & kernel size = 1          \\
                    &                          &                      &                    &                     & stride = 1               \\
                    &                          &                      &                    & Batch normalization  & features = 128           \\
                    &                          &                      &                    &                     & momentum = 0.1           \\
                    &                          &                      &                    & Activation function & ReLu                     \\
                    &                          &                      &                    & Pooling layer       & Adaptive Average pooling \\
                    & \textbf{Fully connected} & features = 128       &                    &                     &                          \\
                    &                          & outputs = 48         &                    &                     &                          \\
                    & \textbf{Batch size = 64} &                      &                    &                     &                          \\
                    & \textbf{Epochs = 30}     &                      &                    &                     &                          \\ \bottomrule
\end{tabular}}
\end{table*}
First, the raw residential load data are averaged to 30 minutes time steps (i.e., 30 minutes granularity). Next, continuous 48h load sequences are constructed using a unit stride sliding window on complete day-long load data. Each one comprises two 24h sequences representing the input and the target load sequences for the forecasting model. Next, time information sequences (i.e., hour of day and day of week) are read from each input load sequence. Then, the resulting sequence data are chronologically split into training and testing sets based on 80/20 ratio (e.g., for load consumption data acquired over 80 days, the first 60 days are set for training the model, and the last 20 for testing it). Subsequently, each load sequence is decomposed to a number of IMFs (and a residue) via VMD (as seen in Fig.\ref{fig:Process}). The ensuing input and forecast sequences from the training and testing sets are standardized to have a zero mean and a unit standard deviation. This is done to simplify the calculation and amplify the forecasting model's convergence speed. Accordingly, the forecasted sequences will have to be reversely standardized to provide the actual predictions. Finally, every resulting decomposed load sequence, and its corresponding time sequences, are fed to a deep learning model as a 3D array of size [number of instances x 3 x 48]. The first dimension represents the training set size. The second dimension represents the number of the input sequences, which reflects a single decomposed load sequence and two time sequences, while, the third dimension represents the length of the input sequences (48 elements). 
Thus, in the context of this paper Eq. (\ref{eq:forecasting}) becomes:
\begin{equation}
 \begin{split}
\hat{E}_{T+S|T+1} = f\big(E_{T-S+1|T},H_{T-S+1|T},D_{T-S+1|T},\epsilon\big)\\, S=48
\end{split}
 \label{eq:forecasting2}
\end{equation}

In all experiments, the hyper-parameters of the VMD algorithm are set as follows. The penalty parameter is 1000, the number of IMFs is $K$, the initial centre frequency is 0, and the convergence criterion is $5\times10^{-6}$. In addition, db4 wavelet is selected as the decomposition wavelet and InceptionTime as the core deep learning model of the proposed model. InceptionTime is a state-of-the-art time-series model that was published in 2019 \cite{InceptionTime2019}. It is an ensemble model comprising five Inception blocks. Each is composed of a series of six Inception modules \cite{Inception2015}, which are deep CNN models, followed by a global average pooling layer and a fully connected layer. In addition, residual connections are implemented between three consecutive Inception modules. The hyper-parameters of the proposed model are listed in Table \ref{tab:ParametersModels}. 

\begin{table*}[htbp]
\centering
\caption{Properties of daily load profiles acquired from the targeted households.}
\label{tab:stats}
\resizebox{\textwidth}{!}{%
\begin{tabular}{llclccccccc}
\toprule
\multicolumn{1}{c}{\multirow{3}{*}[-.5em]{Premises}} & \multicolumn{1}{l}{\multirow{3}{*}[-.5em]{Residents}} & \multicolumn{1}{c}{\multirow{3}{*}[-.5em]{\makecell{Acquisition\\frequency}}} & \multicolumn{1}{l}{\multirow{3}{*}[-.5em]{Start date}} & \multicolumn{1}{c}{\multirow{3}{*}[-.5em]{Duration}} & \multicolumn{6}{c}{Daily consumption}                         \\ \cmidrule(l){6-11} 
\multicolumn{1}{c}{}                          & \multicolumn{1}{c}{}                          & \multicolumn{1}{c}{}                               & \multicolumn{1}{c}{}                            & \multicolumn{1}{c}{}                          & \multicolumn{3}{c}{weekdays}   & \multicolumn{3}{c}{weekends}   \\\cmidrule(l){6-11} 
\multicolumn{1}{c}{}                          & \multicolumn{1}{c}{}                          & \multicolumn{1}{c}{}                               & \multicolumn{1}{c}{}                            & \multicolumn{1}{c}{}                          & mean   & max peak & corr. time & mean   & max peak & corr. time \\ \midrule
House 1                         & Family of three teenagers & 1/5Hz                                              & 1-Oct-2020                                        & 60days                                        & 178.82W & 364.37W  & 19h30min  & 176.55W & 360.23W   & 21h30min   \\
House 2                         & Family of two teenagers & 1/5Hz                                              & 27-Jun-2020                                       & 80days                                        & 220.31W & 301.67W  & 22h00min  & 217.69W & 364.29W   & 16h30min   \\
House 3                         & Family of one older son & 1/5Hz                                              & 28-Jun-2020                                       & 77days                                        & 199.79W & 318.65W  & 19h00min  & 207.53W & 415.14W   & 18h30min   \\
House 4                         & Single adult & 1/5Hz                                              & 12-Nov-2020                                       & 94days                                        & 201.44W & 373.9W   & 6h00min   & 218.5W  & 465.43W   & 14h30min   \\
House 5                         & Elderly couple & 1/5Hz                                              & 10-May-2020                                     & 83days                                        & 188.84W & 256.07W  & 18h30min  & 182.26W  & 275.45W  & 19h30min   \\ \bottomrule
\end{tabular}%
}
\end{table*}

\begin{figure*}[!h]
    \centering
    \subfigure[Diurnal profiles of average 30min granularity electricity consumption with distinctions between weekdays and weekends.]
    {
        \includegraphics[width=0.46\textwidth,trim=30 30 30 30, clip]{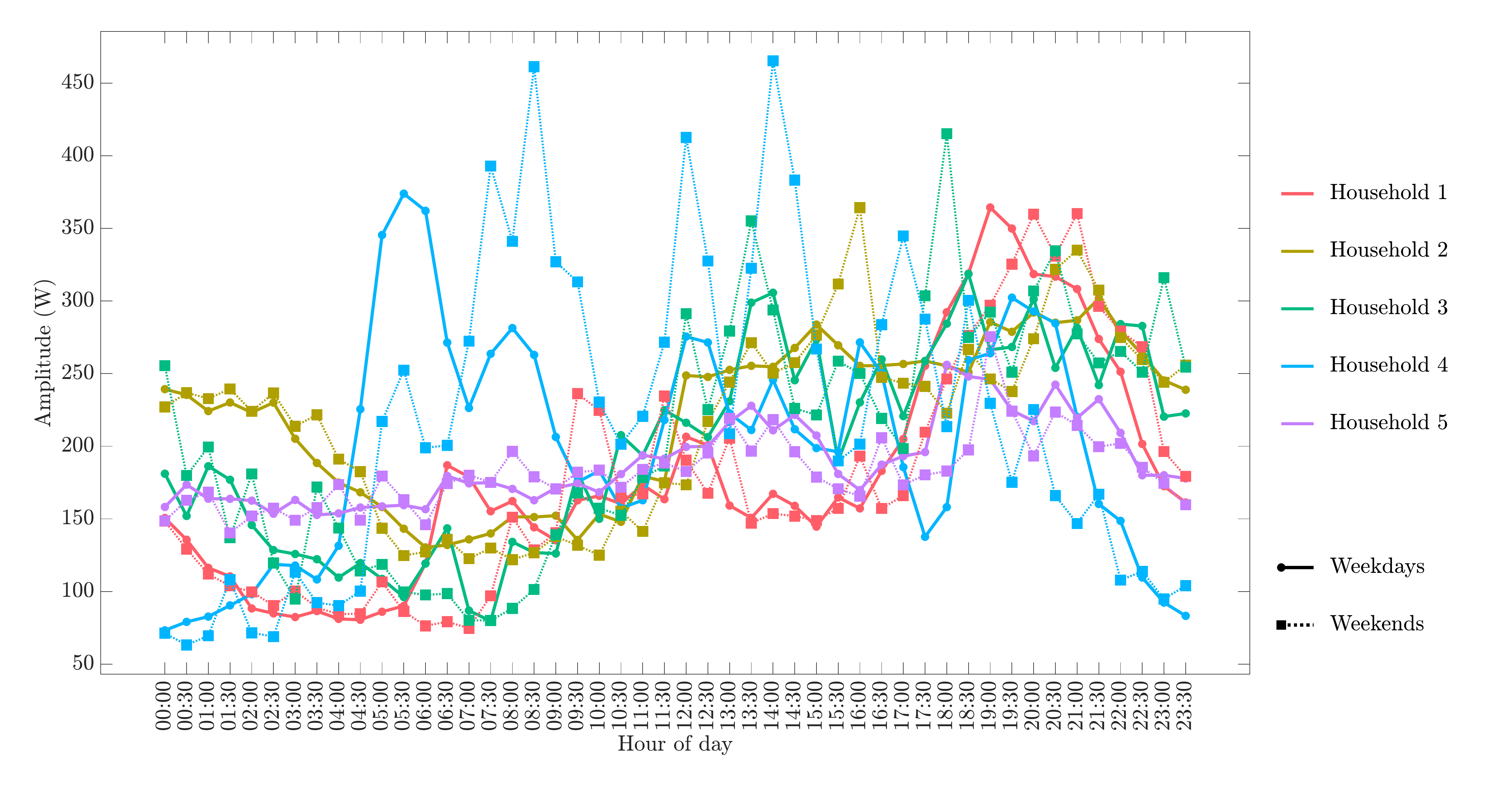}
        \label{fig:AverageLoadHouseholds}
    }
    \
    \subfigure[One-week portions of the electricity consumption profiles.]
    {
        \includegraphics[width=0.46\textwidth]{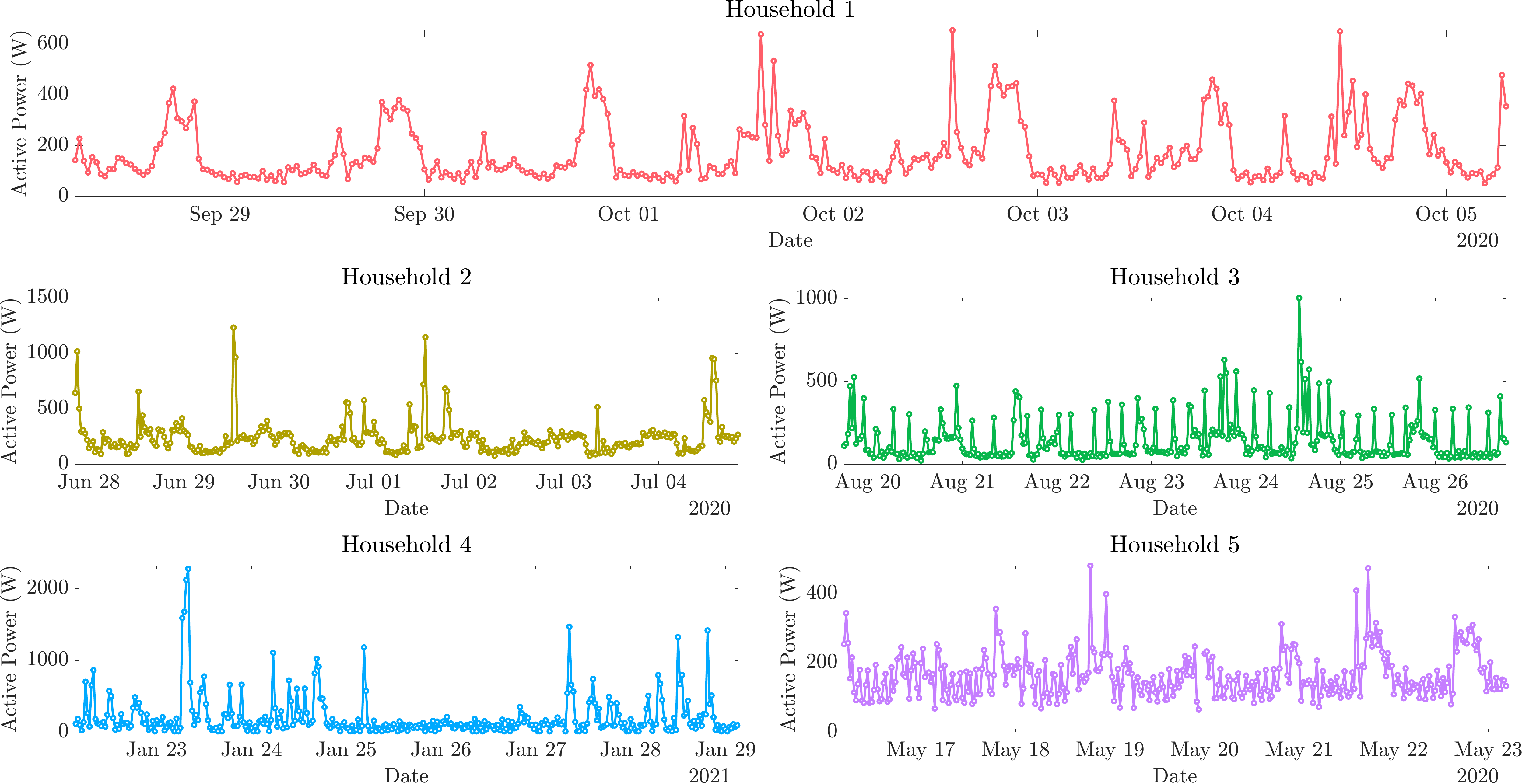}
        \label{fig:Snippets3houses}
    }
    \caption{Electricity consumption profiles acquired from five Moroccan urban households.}
    \label{fig:PropertiesAcquiredData}
\end{figure*}

\subsection{Data acquisition campaign: MORED}
A campaign was launched to collect electricity consumption data from Moroccan urban households. Different types of consumption data were acquired from targeted households, including whole-household consumption data. A number of factors were taken into consideration in order to obtain diverse data, including the type of premises and the socioeconomic status of the neighborhood. The ensuing dataset, coined MORED\cite{MORED}, can be accessed online via this link: \url{https://moredataset.github.io/MORED}. In this paper, the electricity consumption of five mid- to upper-class Moroccan households are utilized. Table \ref{tab:stats} reports some properties of the targeted households.

The targeted households are located on the coast of the country that has a Mediterranean warm climate. Acquisitions were carried out for different durations from summer 2020 to late spring 2021. 
Fig.\ref{fig:PropertiesAcquiredData} showcases examples of load profiles acquired from the five houses. Specifically, Fig.\ref{fig:AverageLoadHouseholds} presents the diurnal patterns of each household averaged over its acquisition duration, while Fig. \ref{fig:Snippets3houses} showcases the corresponding one-week-long raw portions of 30 minutes granularity load profiles. 
In light of these figures, the difficulty encountered in forecasting residential load consumption is readily apparent: load profiles can reflect diverse, complex, and sometimes even volatile consumption patterns. This last characteristic can be seen clearly in household 4 of Fig. \ref{fig:Snippets3houses}, as the corresponding load consumption on the 26th of January 2021 diminishes to approximately half of what it usually is on a weekday. In Fig.\ref{fig:AverageLoadHouseholds}, the average load consumption pattern of house 1 knows a dip in load consumption in the hours following mid-day (i.e., from 13:00 to 18:00), while the opposite is true for the rest of houses, with peak consumption happening at different times (16:00 for household 2 and 13:00 for household 3). Additionally, weekend consumption patterns differ from those observed during the week, as consumption peaks tend to be higher and/or occurring at other intervals of time. Therefore, load and time information are necessary inputs for residential load forecasting.   

\subsection{Performance evaluation}\label{Sec:3.3}
Two main benchmarks are the focus of this paper. The first is regarding the decomposition level and how it affects VMD-powered forecasting. Six discrete values of IMFs (+ Res.) are investigated : 8, 16, 32, 64, 128, and 256. However, the performance of the first two values are not reported in this article as they provide similar or poorer forecasting performance than the baseline model (i.e., historical mean). The second is regarding the proposed model's performance compared with that of other techniques. Additionally, the case of direct forecasting (i.e., no VMD decomposition is utilized) is considered for every model to assess the decomposition step's efficacy in the forecasting task. To this end, a historical mean and four time-series state-of-the-art deep learning models are considered:
\begin{itemize}
    \item Historical mean: 
    A simple baseline technique averaging four consumption features to provide an estimate of the next 24h electricity consumption:
\begin{itemize}
    \item F1: the last 24h consumption of the same-day type (e.g., Last Sunday is the previous same-day type of Saturday, and last Friday is to Monday). 
      \begin{equation*}
F1 =  \left\{\begin{array}{lr}
        E_{T-48|T}, & \text{for Tuesday - Friday}\\
        E_{T-336|T-288}, & \text{for Saturday}\\
        E_{T-192|T-144}, & \text{for Monday}\end{array}\right\}          
      \end{equation*}

    \item F2: the average 24h consumption of the same 24h from the previous three weeks.\\ $F2 = \frac{1}{3}(E_{T-336|T}+E_{T-672|T-336}+E_{T-1008|T-672})$. 
    \item F3: the average 24h consumption of the same 24h from the previous seven days.\\ $F3 = \frac{1}{7}(E_{T-48|T}+E_{T-96|T-48}+E_{T-144|T-96}+E_{T-192|T-144}+E_{T-240|T-192}+E_{T-288|T-240}+E_{T-336|T-288})$. 
    \item F4: the average 24h consumption of the previous day (used as an 48-elements array of the average value).\\ $F4 = \big[\frac{1}{48}E_{T-48|T},\frac{1}{48}E_{T-48|T},...,\frac{1}{48}E_{T-48|T} \big]$.
\end{itemize}
    \item Residual neural network (ResNet)\cite{ResNet2015}:  Introduced in 2015, ResNet has been widely used in computer vision problems. For time-series load forecasting, updated versions of this model are proposed in works such as \cite{Chen2019}. ResNet comprises three stacked residual blocks each containing three convolutional blocks and a linear shortcut linking the block's output with its input to minimize the vanishing gradient effect.       
    \item Omni-Scale 1D-CNN (OS-CNN)\cite{OSCNN2020}: Introduced in 2020, OS-CNN comprises three convolutional layers followed by a global average pooling and a dense layer. To the best of the authors knowledge, OS-CNN has not yet been investigated for the task at hand.
    \item LSTM: Different variations of LSTM-based models have been investigated in the literature for load forecasting \cite{VMDLSTM2019,Transactions2017}. Following \cite{EMDLSTM2020}, a stacked LSTM model is constructed, comprising three LSTM layers of 128 hidden units each, with dropout.    
    \item InceptionTime \cite{InceptionTime2019}:  is a state-of-the-art time-series model that was introduced in 2019. This model's architecture used in the proposed model is used as a separate model as well (refer to Section \ref{Subsection:VMDemporedForecasting} for more details). To the best of the authors knowledge, this model has not yet been investigated for the task at hand.
\end{itemize}

All the considered machine learning models were implemented using the tsai library \cite{tsai} in python, and trained using learning rates, batch sizes, and epochs equal to 0.002, 64, and 30, respectively. The baseline model was developed and assessed using MATLAB.   

The forecasting accuracy of developed models can be evaluated using a variety of metrics. They are usually classified into four categories: absolute errors reflecting the absolute difference between the actual and the predicted value (e.g., Mean Absolute Error (MAE)), percentage errors (e.g., Mean Absolute Percent Error (MAPE)), symmetric errors (e.g., symmetric MAPE), and scaled errors (Mean Absolute Scaled Error).  
To present results comparable with prior studies \cite{AppliedEnergy2021-2,Transactions2017}, the forecasting performance is assessed using four metrics: MAPE, RMSE, coefficient of variance (CV), and forecast skill (FS). These metrics are defined over the whole forecasting horizons following equations \eqref{eq:MAPE}-\eqref{eq:RMSE}:
\begin{equation}
 MAPE = \frac{1}{N\times H}\sum_{i=1}^{N}\sum_{h=1}^{H}\Bigg\lvert\frac{y_{i}(h)-f_{i}(h)}{y_{i}(h)}\Bigg\rvert
 \label{eq:MAPE}
\end{equation}

\begin{equation}
 RMSE = \sqrt{\frac{1}{N\times H}\sum_{i=1}^{N}\sum_{h=1}^{H}\Big(y_{i}(h)-f_{i}(h)\Big)^2}\label{eq:RMSE}
\end{equation}

\begin{equation}
 FS = \Bigg(1-\bigg(\frac{RMSE_{model}}{RMSE_{historic mean}}\bigg)^2\Bigg)\times 100\label{eq:FS}
\end{equation}

\begin{table*}[!h]
\centering
\caption{ADF and KPSS tests ($\alpha=1\%$) on the processed load sequences from the targeted households using either no decomposition or VMD with different levels of decomposition.}
\label{tab:ADF_KPSS_test}
\resizebox{.7\textwidth}{!}{%
\begin{tabular}{llcccccc}
\toprule
\multirow{2}{*}[-.3em]{Household}     & \multicolumn{1}{l}{\multirow{2}{*}[-.3em]{\makecell{Type of\\input sequences}}} & \multicolumn{3}{c}{ADF}                 & \multicolumn{3}{c}{KPSS}                \\ \cmidrule(l){3-5}\cmidrule(l){6-8} 
                               &                        & p-value  & test statistic & critical value & p-value  & test   statistic & critical value \\ \midrule
Household 1                    & no decomposition       & 0.15266 & -1.57454         & \textbf{-2.61480}   & 0.01870 & 0.35178          & \textbf{0.21600}    \\
                               & 31IMFs + Res.          & 0.00081 & \textbf{-0.31119}         & -0.03691   & 0.00120 & \textbf{0.00176}          & 0.00305    \\
                               & 63IMFs + Res.          & 0.00152 & \textbf{-0.62059}         & -0.07382   & 0.00234 & \textbf{0.00398}          & 0.00610    \\
                               & 127IMFs + Res.         & 0.00304 & \textbf{-1.23188}         & -0.14764   & 0.00467 & \textbf{0.00807}          & 0.01220    \\
                               & 255IMFs + Res.         & 0.00670 & \textbf{-2.62389}         & -0.29528   & 0.00930 & \textbf{0.01718}          & 0.02439    \\\addlinespace
Household 2                    & no decomposition       & 0.21206 & -1.37157         & \textbf{-2.61480}   & 0.02379 & 0.35138          & \textbf{0.21600}    \\
                               & 31IMFs + Res.          & 0.00059 & \textbf{-0.23309}         & -0.02755   & 0.00091 & \textbf{0.00124}          & 0.00228    \\
                               & 63IMFs + Res.          & 0.00111 & \textbf{-0.46373}         & -0.05510   & 0.00176 & \textbf{0.00292}          & 0.00455    \\
                               & 127IMFs + Res.         & 0.00223 & \textbf{-0.91938}         & -0.11021   & 0.00350 & \textbf{0.00599}          & 0.00910    \\
                               & 255IMFs + Res.         & 0.00482 & \textbf{-2.00072}         & -0.22041   & 0.00696 & \textbf{0.01256}          & 0.01821    \\\addlinespace
Household 3                    & no decomposition       & 0.00822 & \textbf{-3.04015}         & -2.61480   & 0.05761 & \textbf{0.16320}          & 0.21600    \\
                               & 31IMFs + Res.          & 0.00058 & \textbf{-0.25656}         & -0.02897   & 0.00097 & \textbf{0.00125}          & 0.00239    \\
                               & 63IMFs + Res.          & 0.00104 & \textbf{-0.51397}         & -0.05795   & 0.00190 & \textbf{0.00274}          & 0.00479    \\
                               & 127IMFs + Res.         & 0.00201 & \textbf{-1.05243}         & -0.11589   & 0.00379 & \textbf{0.00546}          & 0.00957    \\
                               & 255IMFs + Res.         & 0.00424 & \textbf{-2.15794}         & -0.23178   & 0.00757 & \textbf{0.01117}          & 0.01915    \\\addlinespace  
Household 4	                   &no decomposition	&0.01026	&\textbf{-3.17669}	&-2.61480	&0.07141	&\textbf{0.12913}	&0.21600\\
	&31IMFs + Res.	&0.00041	&\textbf{-0.15882}	&-0.01936	&0.00065	&\textbf{0.00079}	&0.00160\\
	&63IMFs + Res.	&0.00072	&\textbf{-0.31973}	&-0.03873	&0.00126	&\textbf{0.00180}	&0.00320\\
	&127IMFs + Res.	&0.00141	&\textbf{-0.65906}	&-0.07746	&0.00252	&\textbf{0.00364}	&0.00640\\
	&255IMFs + Res.	&0.00302	&\textbf{-1.39460}	&-0.15492	&0.00501	&\textbf{0.00770}	&0.01280\\\addlinespace
Household 5 & no decomposition & 0.17381 & -1.48082 & \textbf{-2.61480} & 0.05280 & \textbf{0.18605} & 0.21600\\
	&31IMFs + Res.	& 0.00056 & \textbf{-0.23147} & -0.02673 & 0.00089 & \textbf{0.00112} & 0.00221\\
	&63IMFs + Res.	& 0.00101	& \textbf{-0.45911} & -0.05347 & 0.00174 & \textbf{0.00259} & 0.00442\\
	&127IMFs + Res.	& 0.00198	& \textbf{-0.92453} & -0.10693 & 0.00346 & \textbf{0.00522} & 0.00883\\
	&255IMFs + Res.	& 0.00430	& \textbf{-1.93301} & -0.21386 & 0.00687 & \textbf{0.01129} & 0.01767\\\bottomrule
\end{tabular}}
\end{table*}
\begin{equation}
 CV = \frac{\sqrt{\frac{1}{N\times (H-1)}\sum_{i=1}^{N}\sum_{h=1}^{H}\big(y_{i}(h)-f_{i}(h)\big)^2}}{\bar{y}}\times 100\label{eq:CV}
\end{equation}
where $y_i$ is the $h$th actual value of a test set sequence $i$, $f_i$ is the $h$th forecasted value of the same test set sequence, $N$ is the number of test set sequences, $H$ is the number of horizons to be forecasted (i.e., 48), $bar{y}$ is the mean of $y_i$, ${RMSE}_{model}$ and ${RMSE}_{historicmean}$ are the RMSE values of the considered model and the historic mean model resp.  

Results and analysis of these benchmarks using the acquired data are presented in the next section.  

\section{Results}\label{Sec:4}
\subsection{Stationarity analysis}

A stationary time-series exhibits constant statistical properties (e.g., mean) over time. Residential load profiles are generally non-stationary, which makes the forecasting task more challenging. 
Among others, the Augmented Dickey-Fuller (ADF) test \cite{ADF} and the Kwiatkowski, Phillips, Schmidt, and Shin (KPSS) test \cite{KPSS} are two types of tests that can be employed to determine the stationarity of a time-series. 

ADF assesses the null hypothesis that a time-series possesses a unit root against the alternative hypothesis that it is stationary (or stationary around a deterministic trend (trend-stationary)). 
The corresponding test statistic value must be lower than the critical value to reject the null hypothesis and hence prove the alternative. 
While KPSS assesses the null hypothesis that a time-series is trend-stationary against the alternative that it is a unit root non-stationary process. 
The corresponding test statistic value must be higher than the critical value to reject the null hypothesis and prove the alternative.

Table \ref{tab:ADF_KPSS_test} presents the results of ADF and KPSS tests run on the pre-processed load sequences (i.e., input data to the forecasting model) from target households. Tests were run using sequences without decomposition as well as VMD-decomposed sequences to 31, 63, 127, or 255 IMFs in order to assess the effectiveness of VMD. Based on the table, it can be seen that both tests confirm that the VMD-decomposed sequences are stationary, which is not always the case otherwise. In fact, in the ADF test, test statistics for sequences with no decomposition are higher than their corresponding critical values in most households data (i.e., households 1, 2, and 5). The opposite is found when they are decomposed. Similarly, in the KPSS test, test statistics are lower than critical values with decomposed sequences. In addition, it can be seen that the higher the level of decomposition, the more significant the difference between the two values is (i.e., the test statistic and the critical value), which proves that deeper decomposition levels result in sequences with more substantial proof of stationarity. 
\begin{figure*}[!t]
    \centering
    \includegraphics[width=0.85\textwidth]{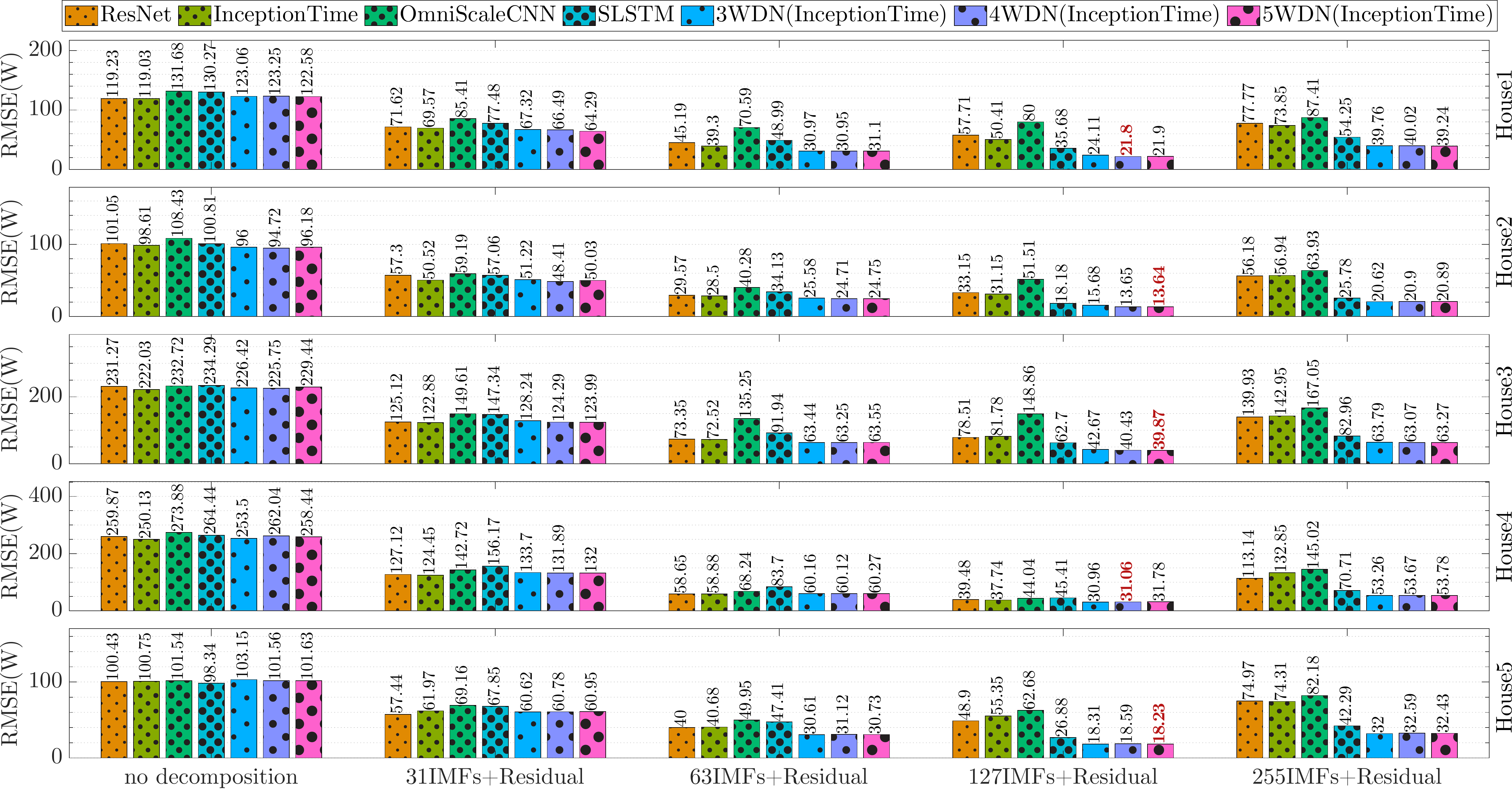}
    \caption{The forecasting performance of all considered models assessed using RMSE and all households (best values are red bold).}
    \label{fig:ForecastingPerformance_RMSE}
\end{figure*}

\begin{figure*}[!t]
    \centering
    \includegraphics[width=0.85\textwidth]{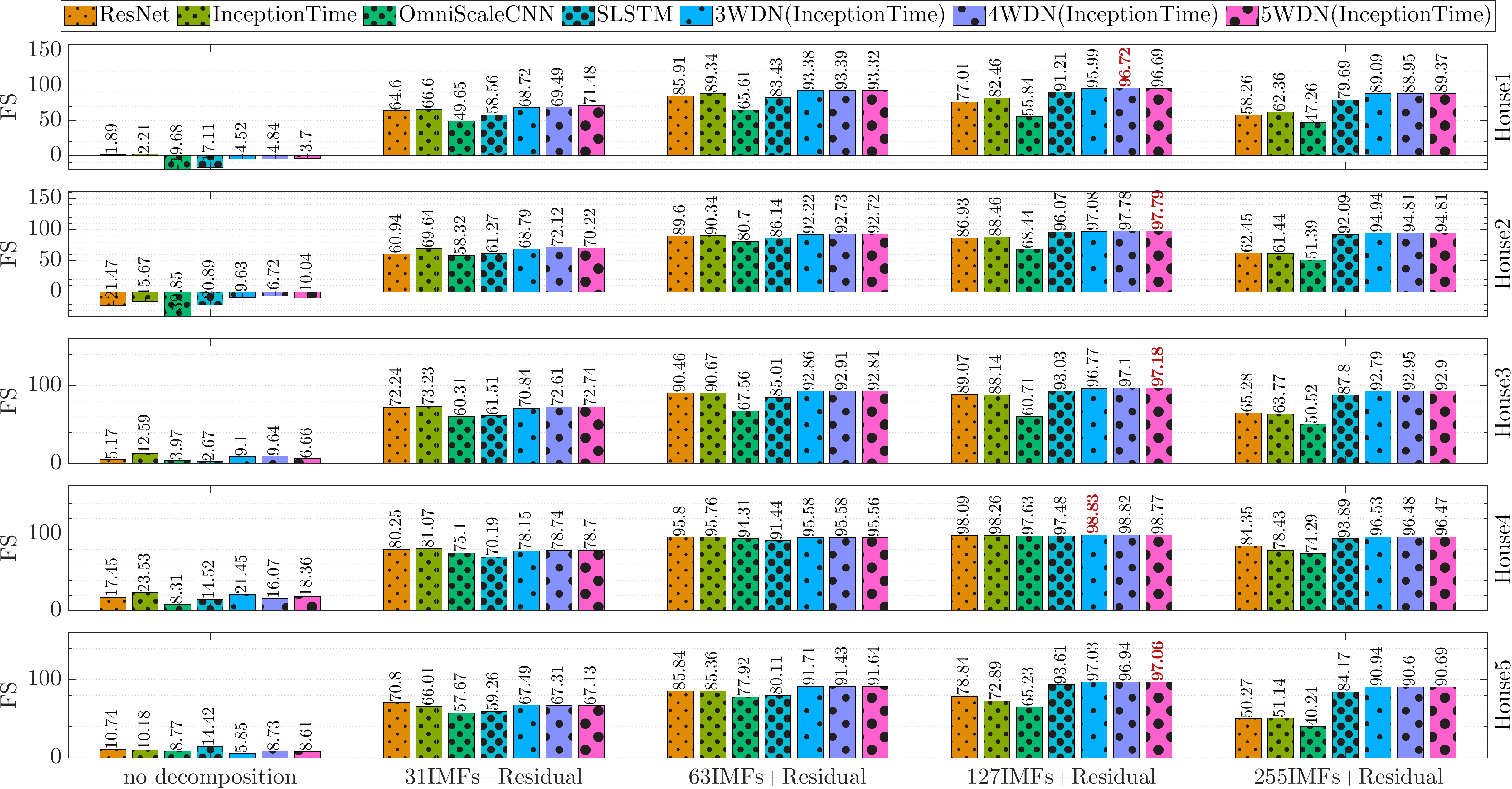}
    \caption{The forecasting performance of all considered models assessed using FS and all households (best values are red bold).}
    \label{fig:ForecastingPerformance_FS}
\end{figure*}

\subsection{Best decomposition level and best performing model}

\begin{figure*}[!t]
    \centering
    \includegraphics[width=0.85\textwidth]{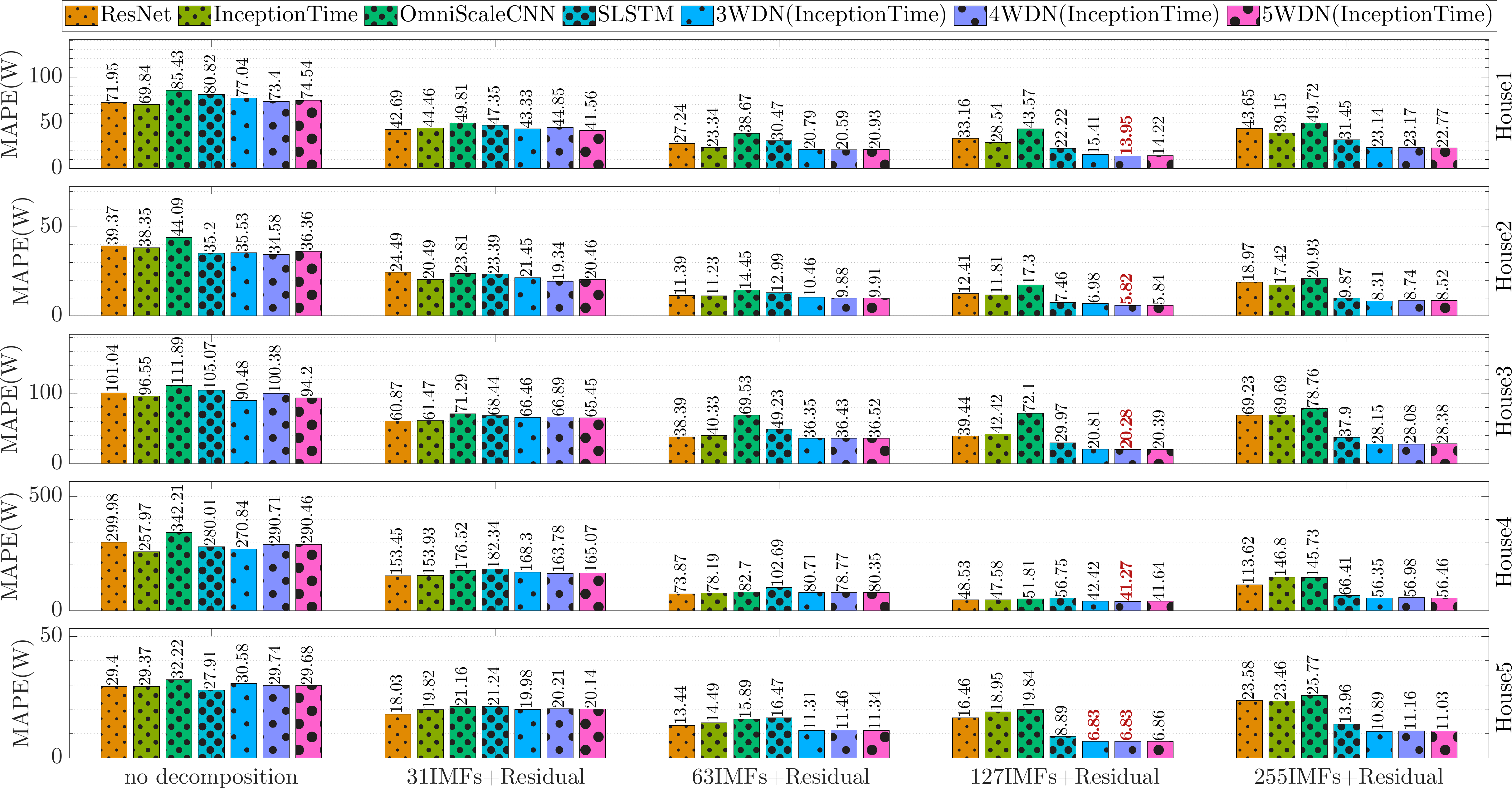}
    \caption{The forecasting performance of all considered models assessed using MAPE and all households (best values are red bold).}
    \label{fig:ForecastingPerformance_MAPE}
\end{figure*}

The prior detailed experiments were carried out on the acquired whole-house energy consumption of five houses. Table \ref{tab:HistoricalMeanForecast} shows the forecasting performance of the historical mean model on all five houses. Figs. \ref{fig:ForecastingPerformance_RMSE}-\ref{fig:ForecastingPerformance_MAPE} present the values of RMSE, FS, CV, and MAPE of the proposed model (i.e., mWDN with InceptionTime as its base model) and the considered benchmark models for all targeted houses, respectively. Three variations of the proposed model (3, 4, and 5 levels of wavelet decomposition) are considered to identify the optimal model for the task at hand. The following observations can be gathered from the reported performances:

\begin{figure*}[!t]
    \centering
    \includegraphics[width=0.85\textwidth]{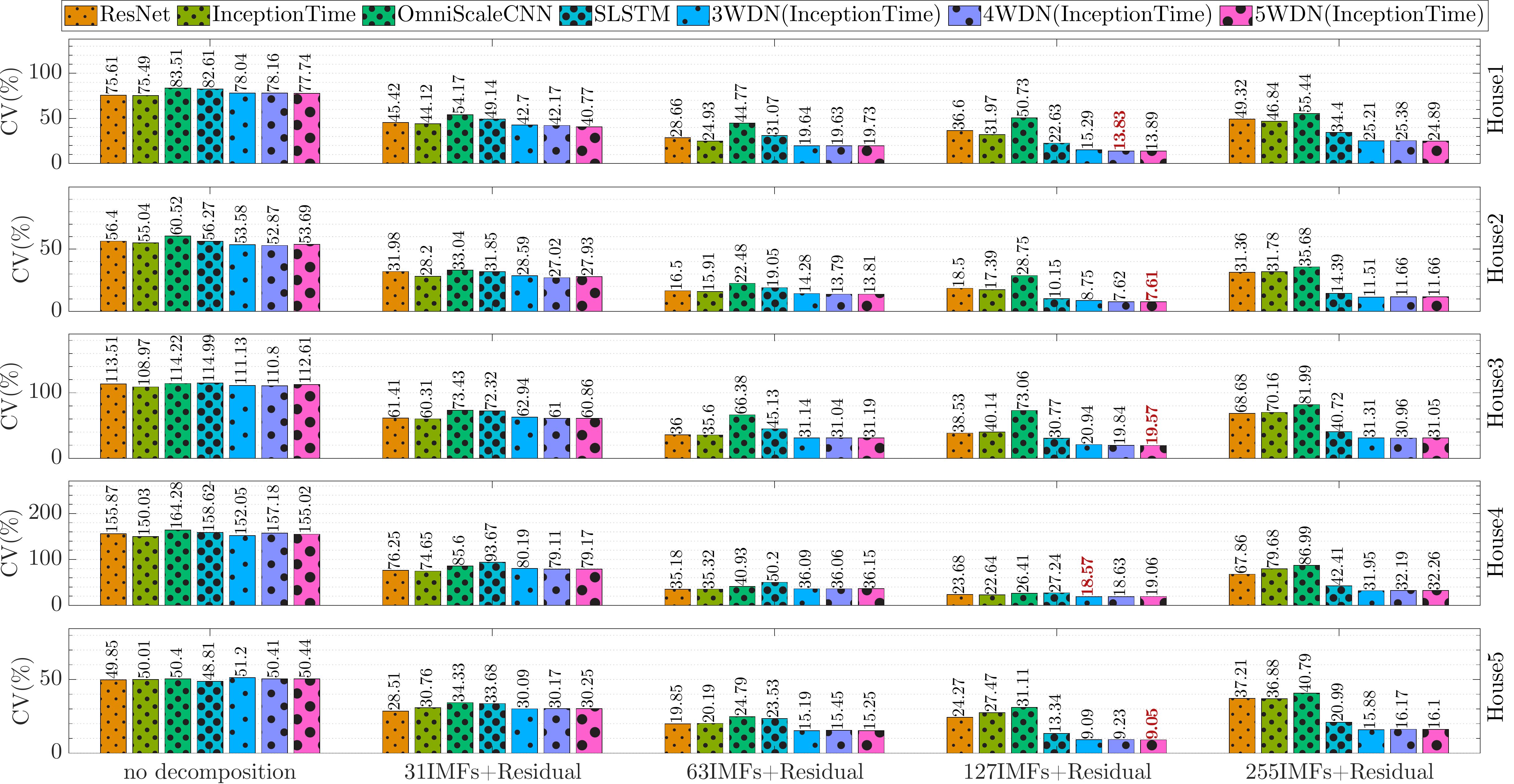}
    \caption{Forecasting performance of models using CV on all households (best values are red bold).}
    \label{fig:ForecastingPerformance_CV}
\end{figure*}

\begin{itemize}
    \item In the absence of VMD, all models perform poorly. OmniScale-CNN consistently yields the poorest performance in terms of MAPE in all households (85.43W, 44.09W, 111.89W, 342.21W, and 32.22W resp.) and in terms of the rest of the considered metrics in houses 1, 2, and 4. SLSTM follows it in terms of RMSE and FS for houses 1 (130.27W and 7.11\% resp.) and 4 (264.44W and 14.52\% resp.), and ResNet in terms of RMSE and CV in houses 2 (101.05W and 56.4\% resp.) and in terms of MAPE in both houses 2 and 4 (38.35W and 299.98W resp.). While the mWDN(InceptionTime) variations provide the poorest performances in terms of RMSE, FS, and CV only in house 5. Furthermore, the historical mean model achieves slightly better performances than the other models only in houses 1 and 2 in terms of all metrics. Higher performances are achieved in the rest of the houses using the other models. This demonstrates the advantages of deep learning models in forecasting complex and varying time-series over multiple horizons.
    
 \begin{table}[htbp]
\centering
\caption{Historical mean model forecasting performance on targeted households. }
\label{tab:HistoricalMeanForecast}
\begin{tabular}{lccc}
\toprule
Premises & RMSE(W)    & CV(\%)    & MAPE(W)      \\ \midrule
House1   & 120.373 & 76.339   & 68.81  \\
House2   & 91.689  & 51.177  &  26.002 \\
House3   & 237.484 & 116.634 & 111.755 \\
House4   & 286.026 & 171.564 &  317.211\\
House5   & 106.304 & 52.764  &  32.584 \\ \bottomrule
\end{tabular}%
\end{table}

\begin{figure*}[!t]
    \centering
    \subfigure[House 1]
    {
        \includegraphics[width=0.75\textwidth]{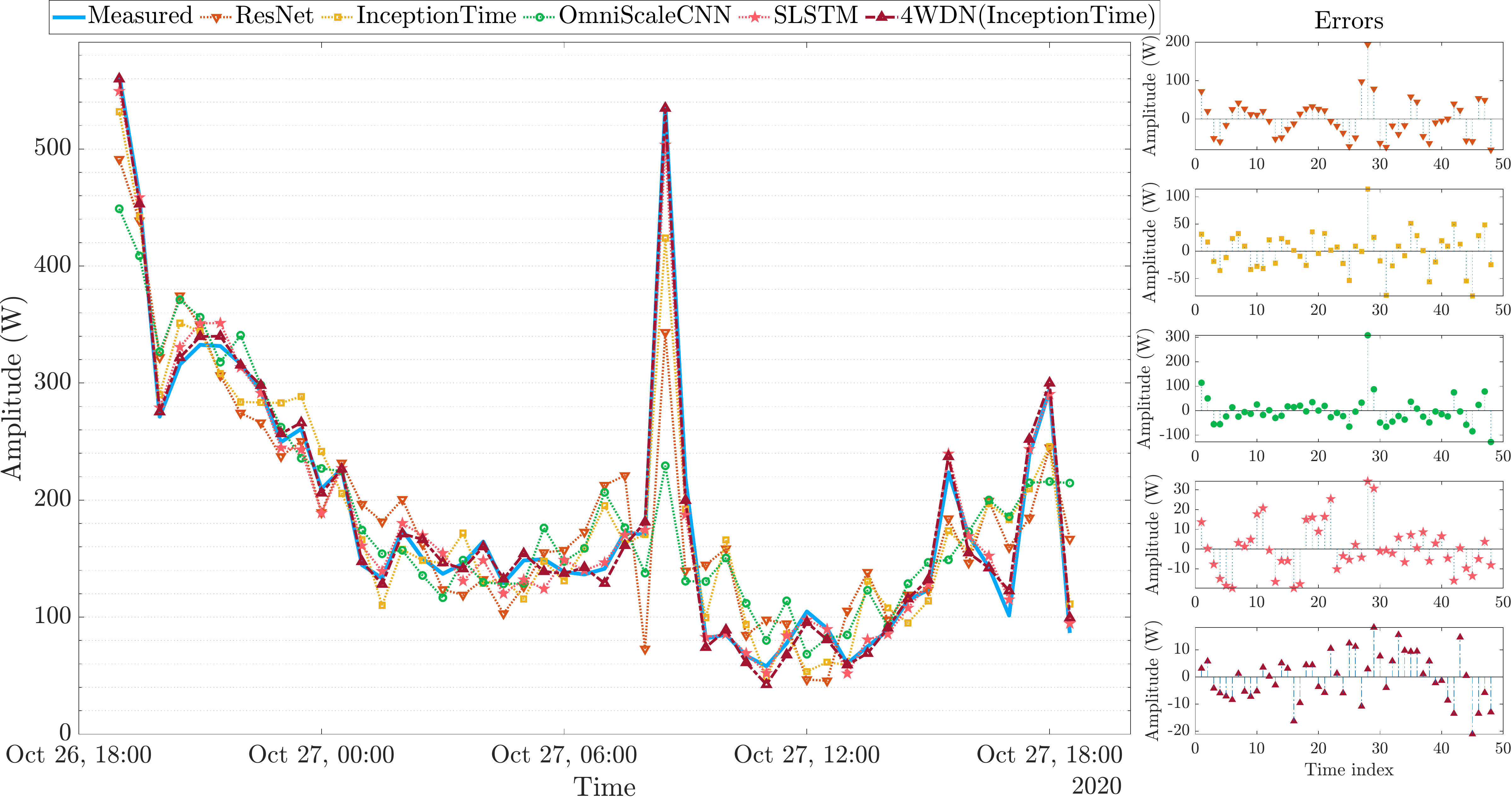}
        \label{fig:House1_forecastingVMD}
    }
    \\
    \subfigure[House 2]
    {
        \includegraphics[width=0.99\columnwidth]{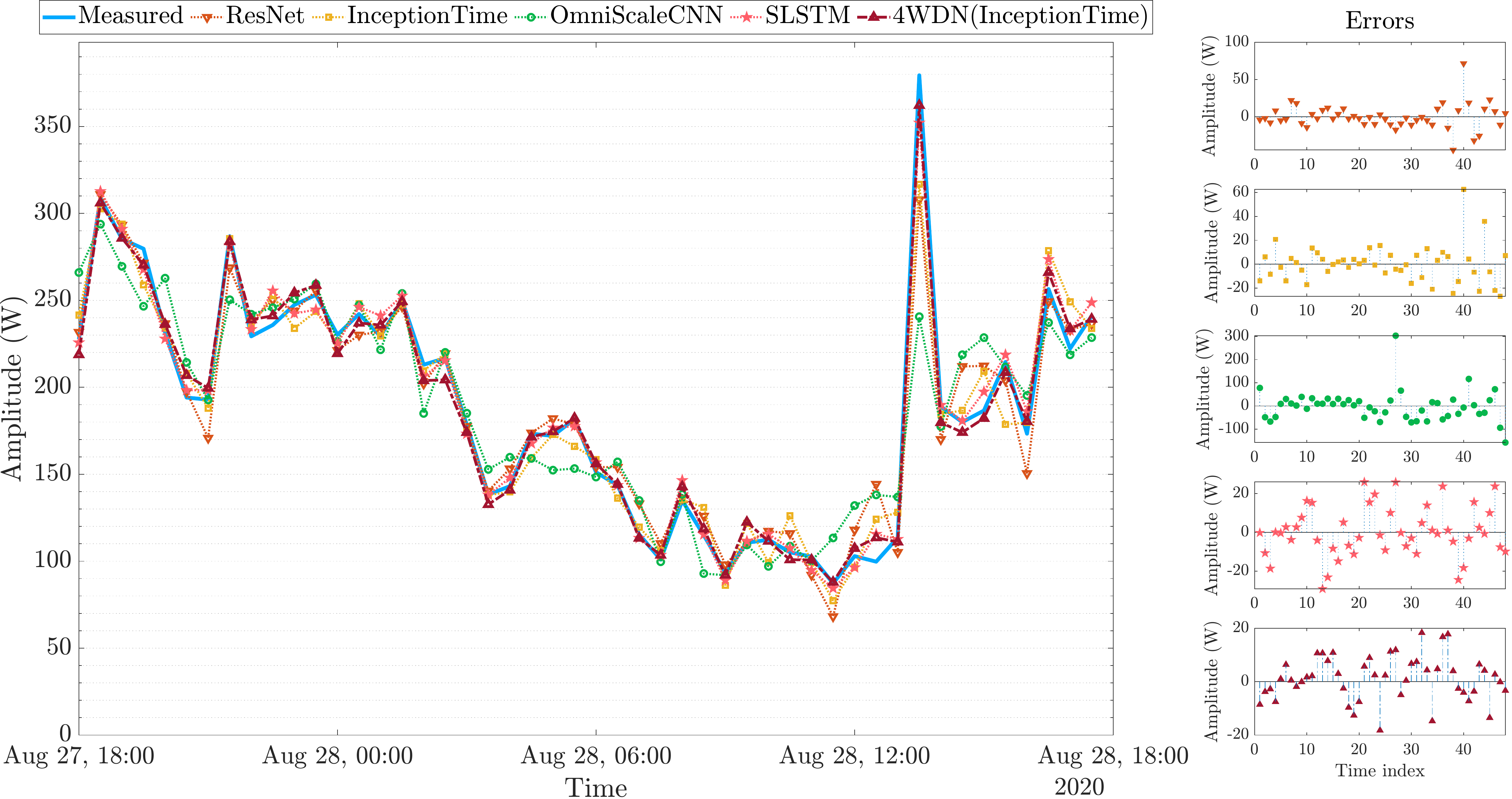}
        \label{fig:House2_forecastingVMD}
    }
    \quad
    \subfigure[House 3]
    {
        \includegraphics[width=0.99\columnwidth]{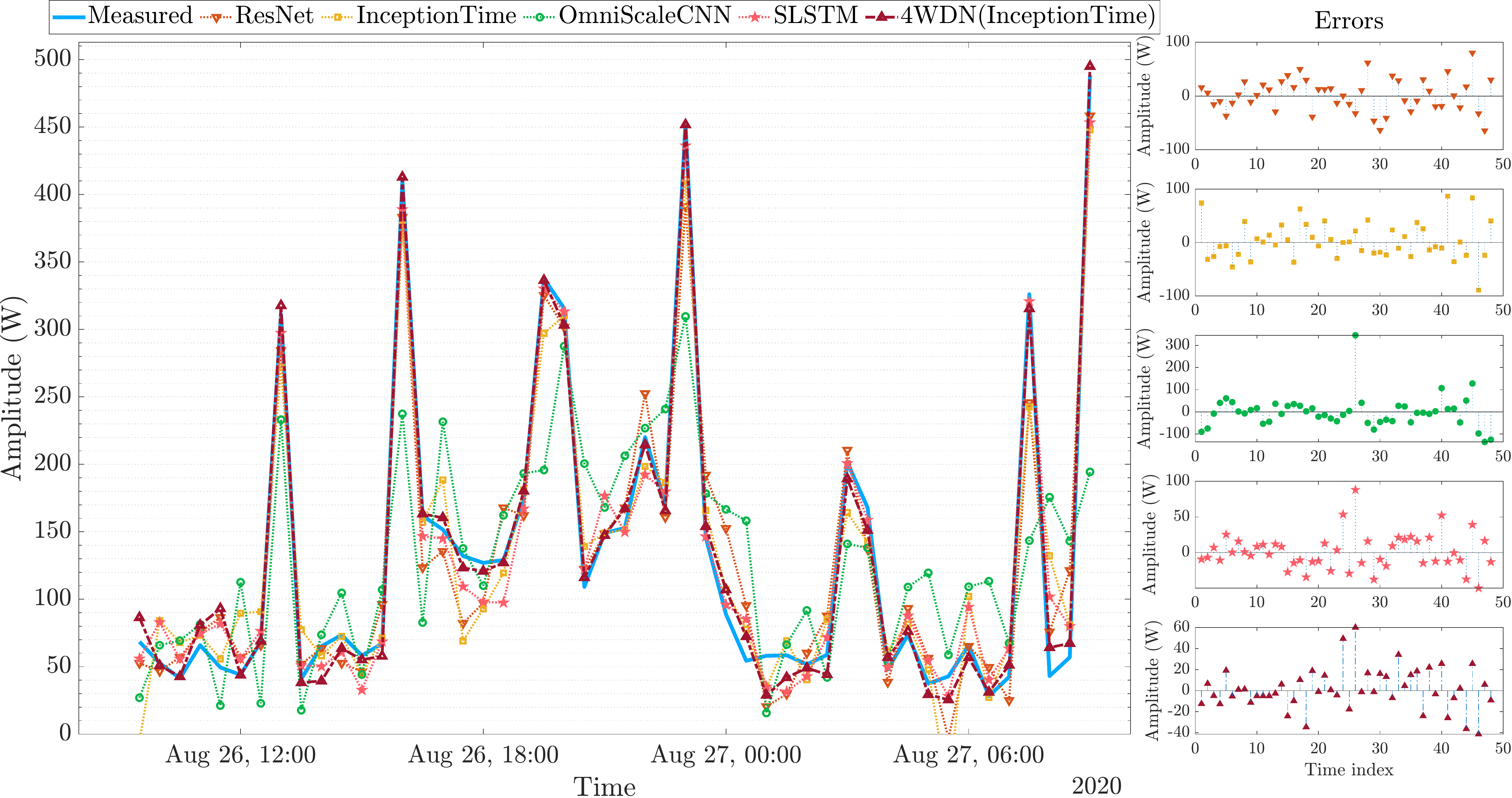}
        \label{fig:House3_forecastingVMD}
    }
    \\
    \subfigure[House 4]
    {
        \includegraphics[width=0.99\columnwidth]{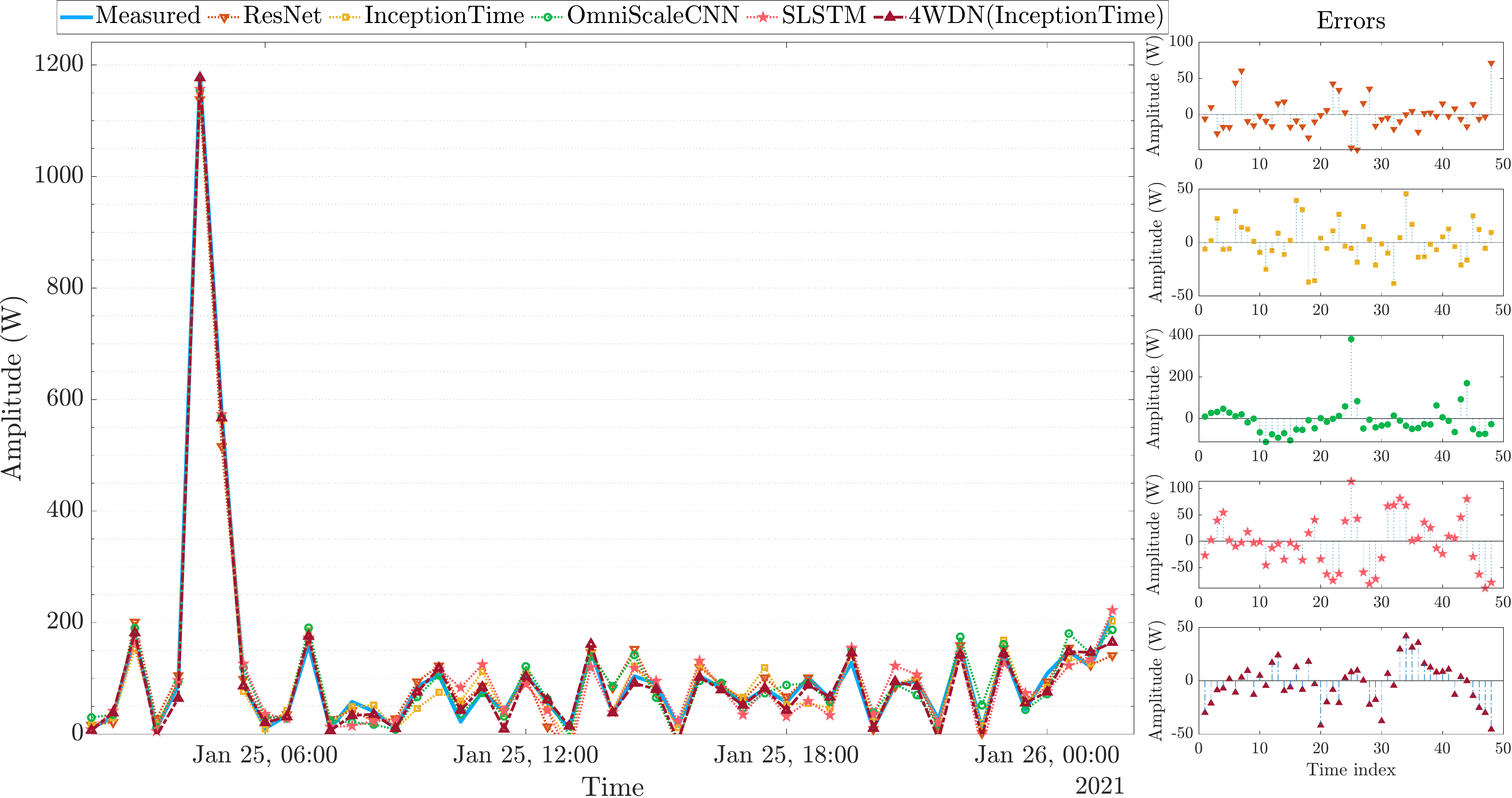}
        \label{fig:House4_forecastingVMD}
    }
    \quad
    \subfigure[House 5]
    {
        \includegraphics[width=0.99\columnwidth]{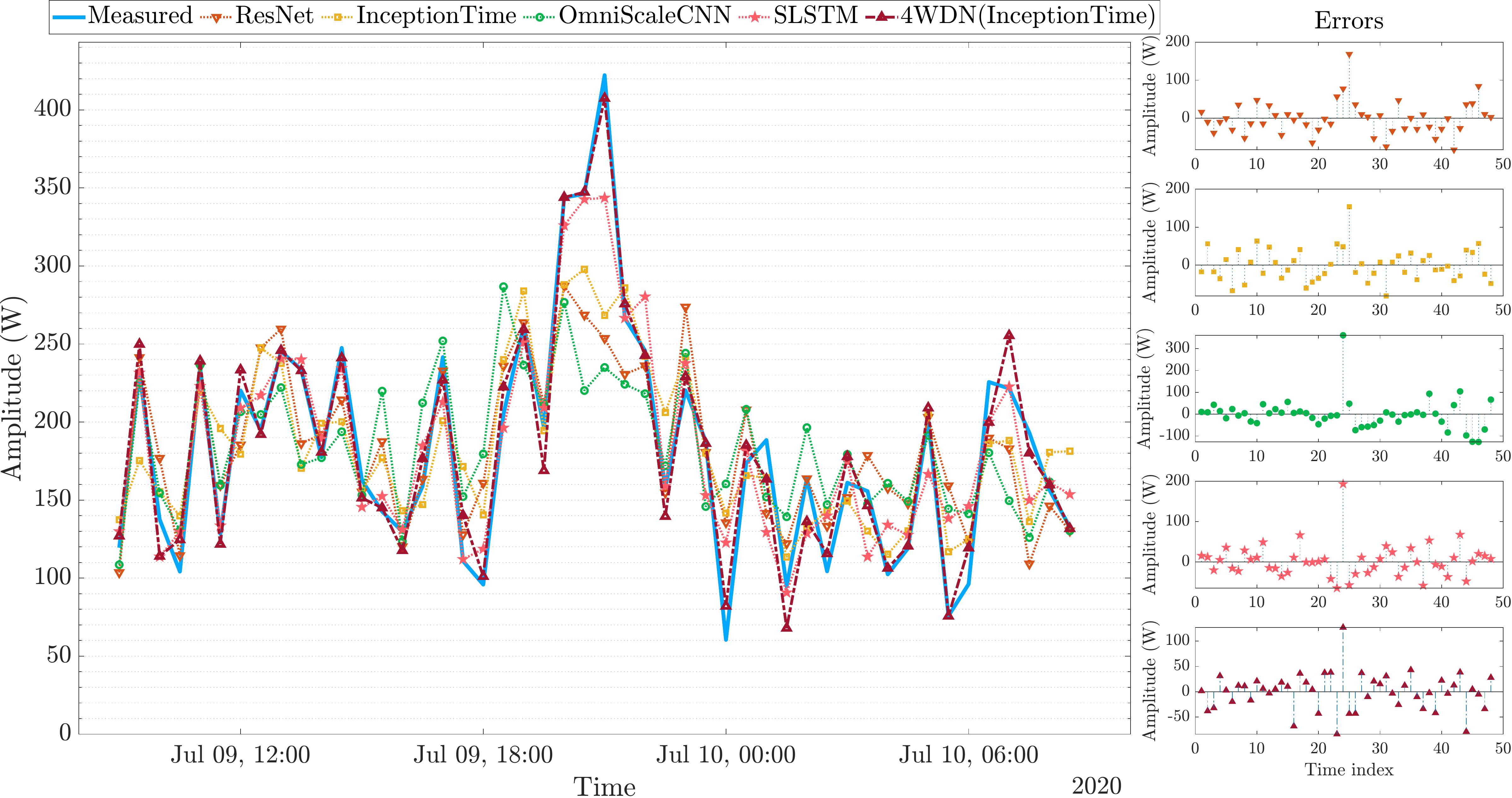}
        \label{fig:House5_forecastingVMD}
    }    
    \caption{A performance comparison of VMD-empowered ($K = 127$) deep learning models forecasting a 24h sequence from the test set of all households.}
    \label{fig:ForecastingVMD_plots}
\end{figure*}

    \item When VMD is incorporated into forecasting techniques, greater accuracy is achieved with higher levels of decomposition, until a certain level. Indeed, ResNet, InceptionTime, and OmniScale-CNN achieve their highest performances across all metrics at $K=63$ for all households except house 4. For instance, these models attain RMSE values equal to 29.57W, 28.5W, and 40.28W, respectively, for house 2 or an improvement of 70.74\%, 71.10\%, and 62.85\%, respectively, from the first case (i.e., no VMD). Their performances decrease with the subsequent decomposition levels (e.g., 33.15W, 31.15W, and 51.51W, respectively, in terms of RMSE at $K=127$ for house 2). A similar trend can also be seen with SLSTM and the mWDN(InceptionTime) variations. However, the highest performances are achieved at a later decomposition level of $K=127$ (e.g., 18.18W, 15.68W, 13.65W, 13.64W, respectively, in terms of RMSE for house 2, or an improvement of 81.97\%, 83.67\%, 85.59\%, 85.82\%, respectively, from the first case). For house 4, all models can achieve their peak performance at $K=127$ in terms of all metrics.        
    \item The top-performing model for all households is mWDN(InceptionTime). It can be clearly seen that the three variations of this model, corresponding to each wavelet decomposition level ($I=3$, $4$, and $5$), consistently provide the top three performances when compared with all the other models at $K=127$ (e.g., 15.68W, 13.65W, 13.64W, respectively, in terms of RMSE for house 2). For houses 1 and 4, the model with a wavelet decomposition level of $I=4$ provides the highest performance in terms of all metrics with an approximate improvement of 82.31\% and 88.14\%, respectively, in terms of RMSE and CV, and 81.00\% and 85.8\%, respectively, in terms of MAPE. For houses 2 and 3, mWDN(InceptionTime) with $I=5$ achieves slightly higher performance than that with $I=4$ in terms of all metrics except MAPE. For house 5, the model with a wavelet decomposition level of $I=3$ achieves the highest performance in terms of all metrics with an improvement of 82.25\% in terms of RMSE, 82.24\% in terms of CV, and 77.66\% in terms of MAPE. 
\end{itemize}

\begin{figure*}[!t]
    \centering
    \subfigure[House 1]
    {
        \includegraphics[width=0.65\textwidth]{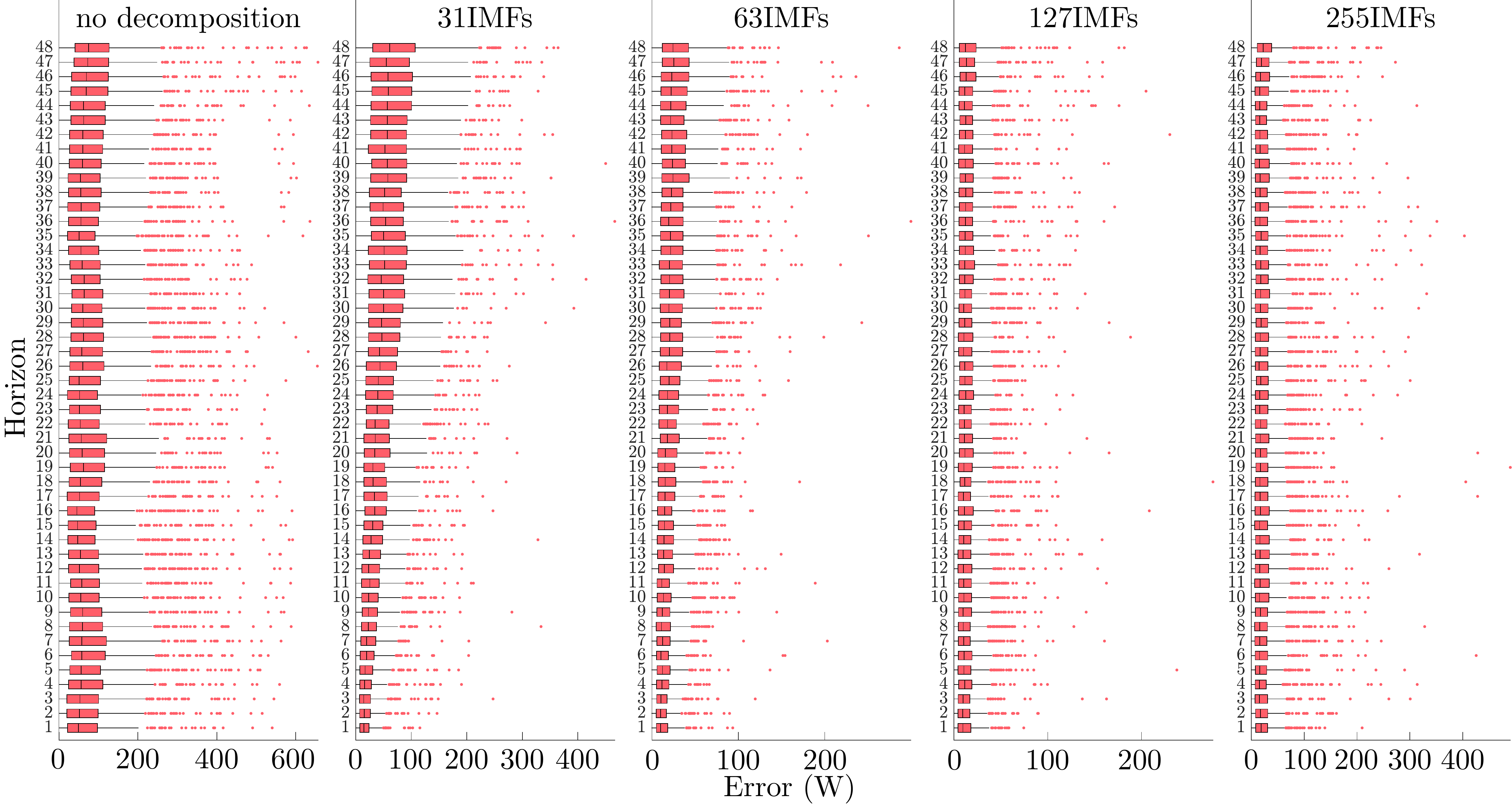}
        \label{fig:House1_boxplot}
    }
    \\
    \subfigure[House 2]
    {
        \includegraphics[width=0.99\columnwidth]{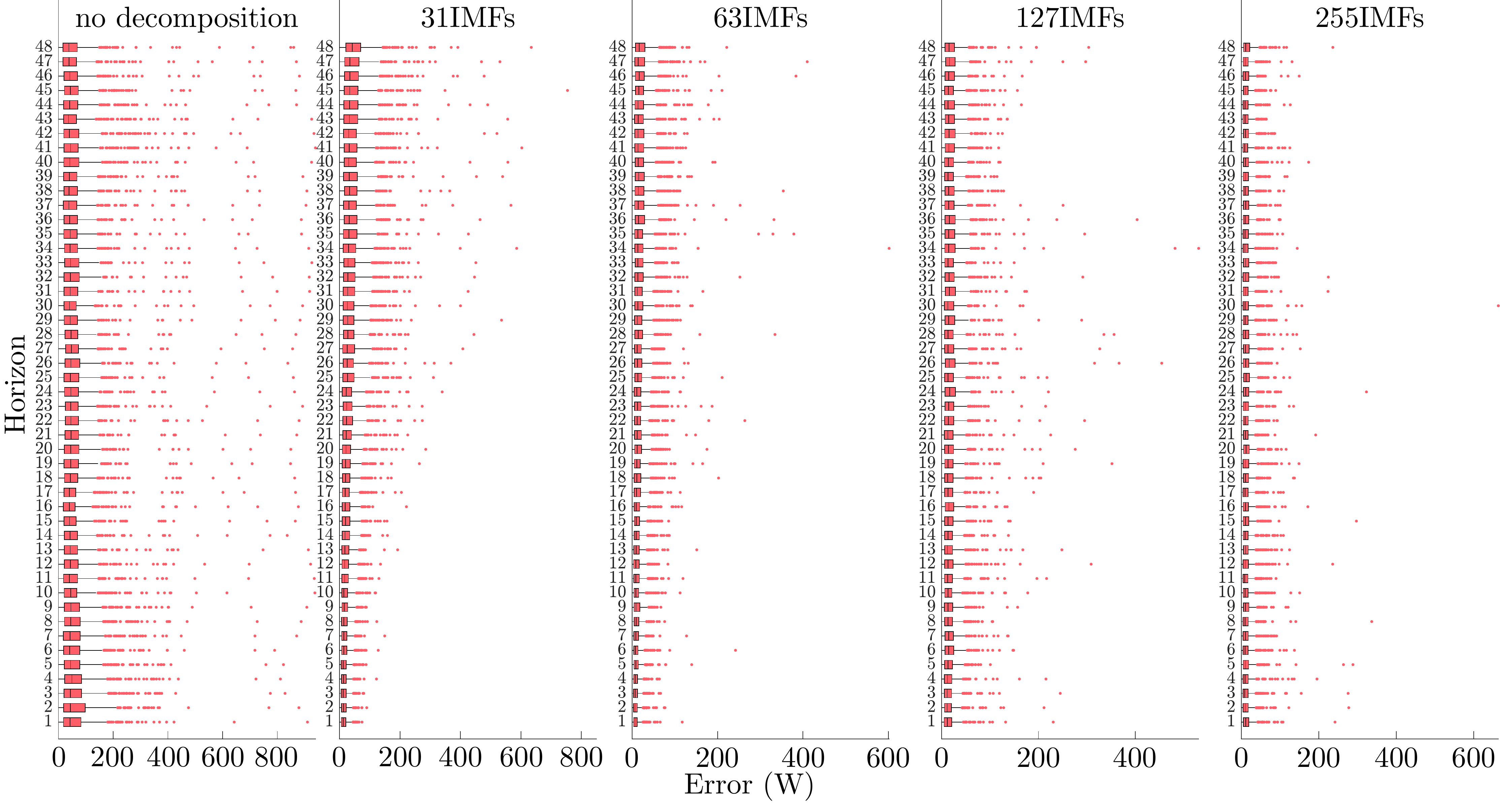}
        \label{fig:House2_boxplot}
    }
    \quad
    \subfigure[House 3]
    {
        \includegraphics[width=0.99\columnwidth]{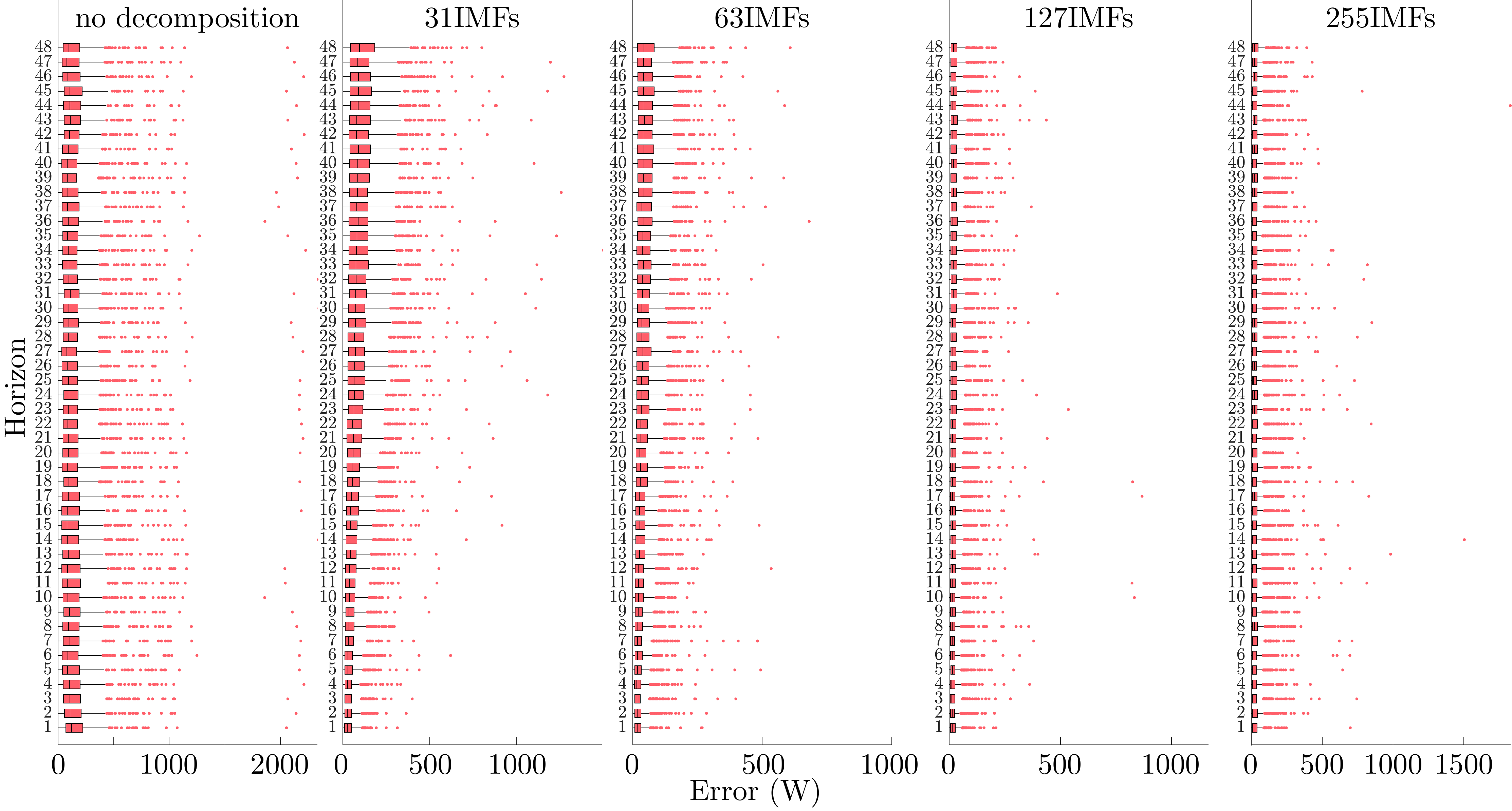}
        \label{fig:House3_boxplot}
    }
    \\
    \subfigure[House 4]
    {
        \includegraphics[width=0.99\columnwidth]{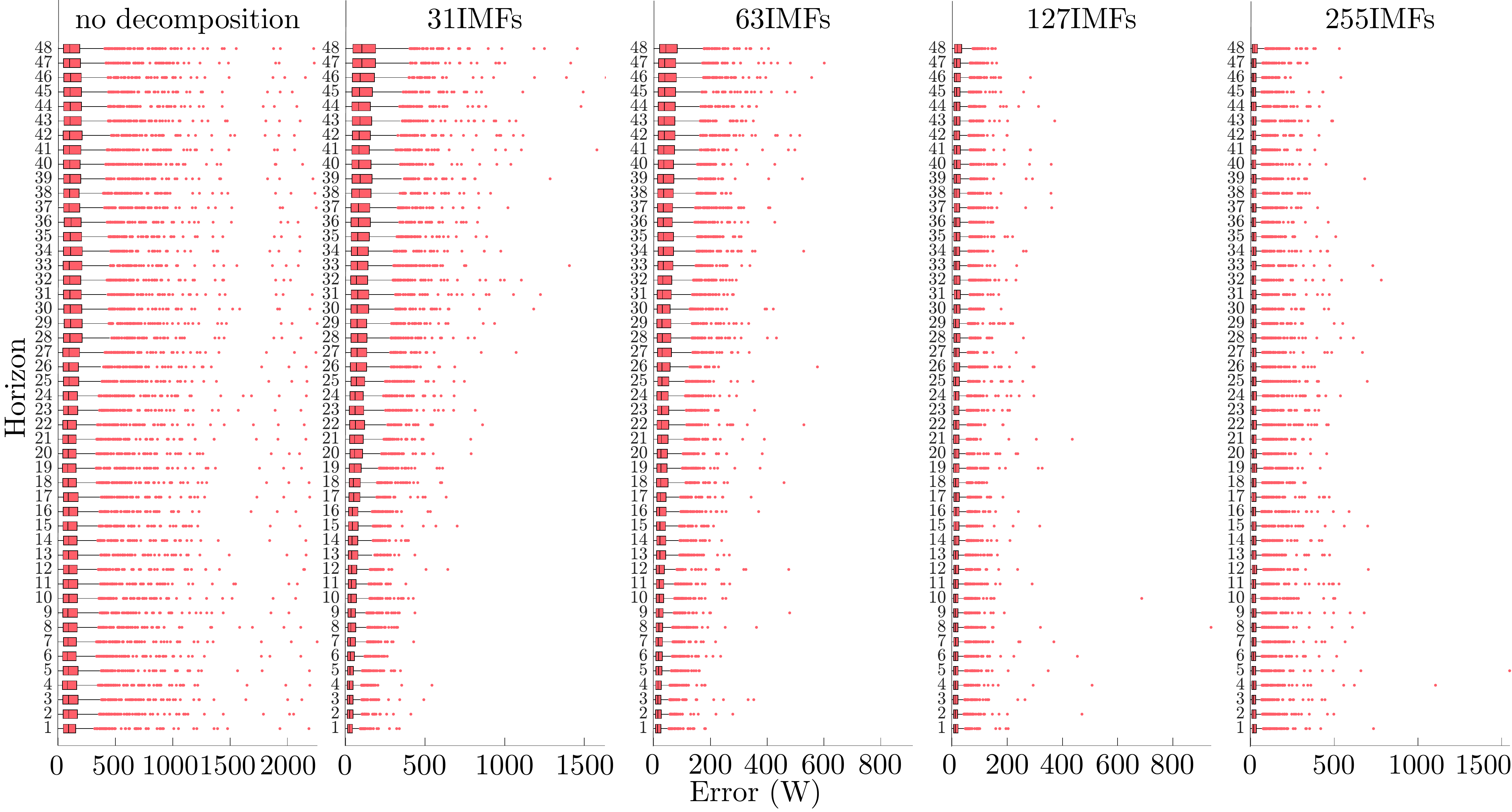}
        \label{fig:House4_boxplot}    
    }
    \quad
    \subfigure[House 5]
    {
        \includegraphics[width=0.99\columnwidth]{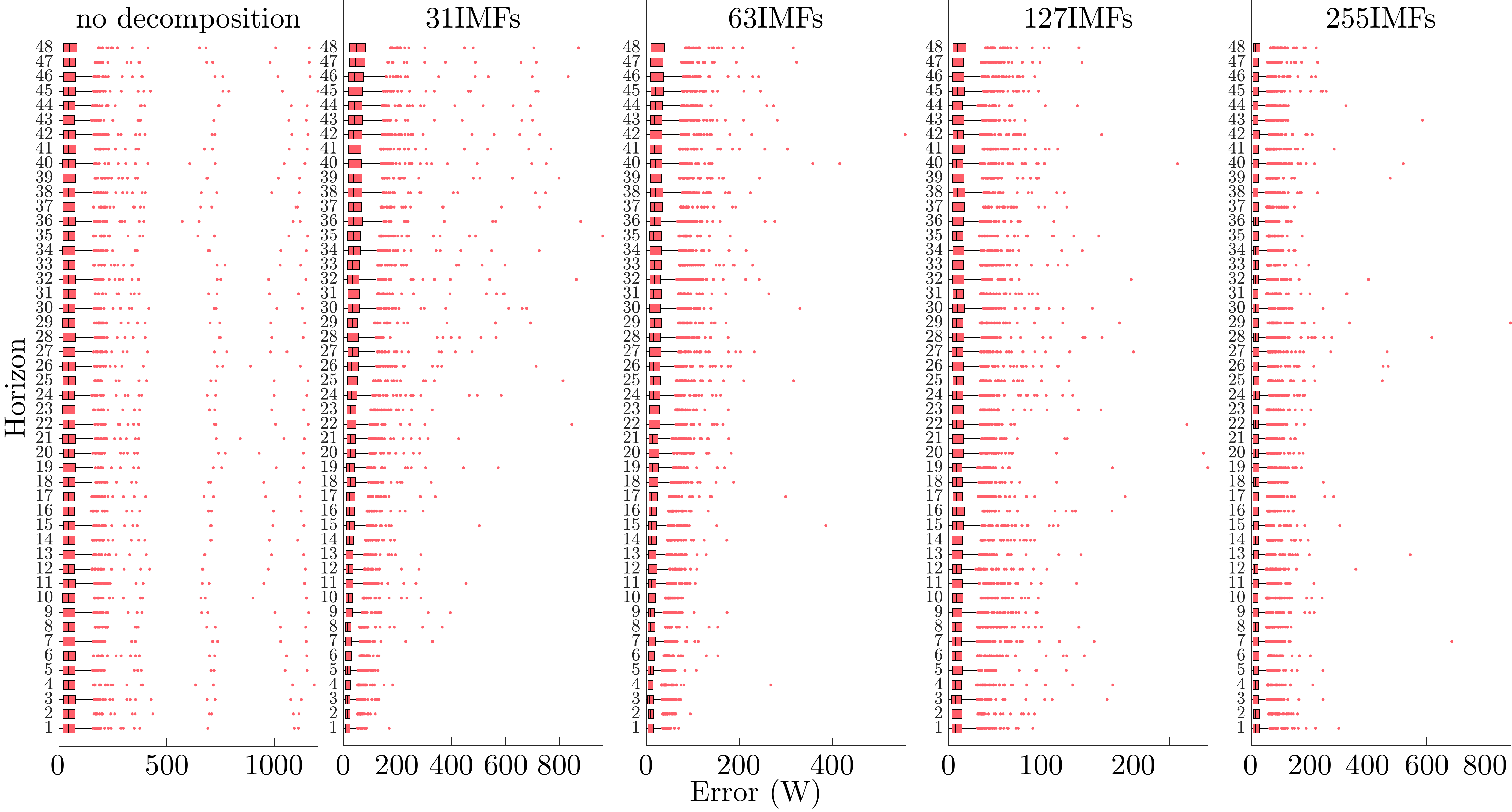}
        \label{fig:House5_boxplot}    
    }    
    \caption{Boxplot of the observed errors along the forecast horizons, averaged over the whole test set of each household using the proposed 4WDN(InceptionTime) model with no decomposition, $K = 31$, $K = 63$, $K = 127$, and $K = 255$.}
    \label{fig:Boxplot}
\end{figure*}

Fig.\ref{fig:ForecastingVMD_plots} illustrates an example of each household's testing set in which the actual and the forecasted load consumption are plotted at all horizons, along with their corresponding errors. For better visualization, only the highest performing VMD-empowered ($K = 127$) models were plotted, where different consumption patterns can be seen for each of the houses. Among all benchmarking models, SLSTM better follows the actual load pattern of all houses, followed by InceptionTime. However, the proposed model outperforms SLSTM in that it is able to forecast the sudden and high peaks in load consumption with greater accuracy. A feat only this model can sustain with very low errors in all households. The reason is that the model makes use of the wavelet decomposition embedded with the base model (i.e., InceptionTime) to extract multi-resolution time-frequency information that is vital to learning various consumption behaviors in the load profile.

Further support for this observation can also be found in Fig.\ref{fig:Boxplot}, which presents the error distribution of only the 4WDN(InceptionTime) model over the whole 24h forecasting duration (i.e., 48 steps) averaged over the whole testing set of each household. In this figure, the model gives the highest errors for all considered horizons when no decomposition is used. Processing the inputs via VMD before injecting them into a forecasting model is imperative to enhance performance as the related errors are seen to decrease throughout the studied scenarios (i.e., no decomposition, $K = 31$, $K = 63$, $K = 127$, and $K = 255$). For these VMD-empowered cases, however, further horizons are associated with increasing errors. Lastly, the most optimal VMD decomposition level of all households resides in the interval $K \in [127,255[$.

\section{Discussion}
In this paper, the developed models are trained and tested on load profiles from five different Moroccan households. These profiles were acquired for periods ranging between two and three months approximately. These acquisitions fall generally within a single season (Summer for houses 2 and 3) or between the start of two seasons (Autumn - Winter for houses 1 and 4, and Spring - Summer for house 5). Considering this low diversity in terms of season in this dataset, a random selection for the training data would not present a significant impact on the performance in our particular case against the use of the deterministic chronological splitting implemented. This was confirmed by initial preliminary tests performed on this dataset. For larger datasets with richer diversity in terms of seasons, the random selection for the training set would be a more appropriate choice.

This study investigated the impact of six decomposition levels of VMD ($K= 8, 16, 32, 64, 128,256$, see section \ref{Sec:3.3}) and three decomposition levels of the wavelet decomposition technique within the mWDN model ($I=3,4,5$), see Figs. \ref{fig:ForecastingPerformance_RMSE}-\ref{fig:ForecastingPerformance_MAPE}. The considered VMD decomposition levels were selected based on earlier studies as well as from multiple preliminary tests. These tests were conducted using different decomposition levels and load profiles. VMD decomposition levels of $K<8$ provided little improvement to the forecasting in comparison to the case where no decomposition was used. Improvements started to be noticed for decomposition levels $K\ge8$. It was also observed that small linear increases on the decomposition level (e.g., $K = K+1$) showed little change in the forecasting performance. Consequently, the decomposition levels of VMD were selected according to an exponential rule that allowed to observe larger changes in the forecasting performance.

Moreover, enhanced forecasting performances were achieved using the proposed model with a VMD decomposition level of $K = 127$. This result was consistently achieved across all considered decomposition levels, models, and houses. These results highlight the importance of incorporating decomposition techniques within forecasting approaches, which provide an alternative approach to relying on exogenous features (e.g., temperature). Indeed, these techniques are especially useful to achieve high forecasting accuracy in communities and households where such features are not particularly influential on electricity consumption throughout the year (due to generally stable weather conditions, financial reasons, etc.). Nevertheless, despite its diverse load profiles and consumption patterns, this dataset can only be regarded as a general indication of how electricity is consumed in similar Moroccan households and conditions. Hence, further investigation is necessary to identify the optimal parameters and forecasting models appropriate with multiple datasets reflecting different communities and conditions. Nevertheless, the authors hope that the present work can be a useful guideline to such future studies.

The practical implementation of the proposed technique in an online forecasting setting is feasible. Indeed, a sliding window of 48 elements in width and a unit stride will be able to provide the necessary load and time sequences sliding every 30 minutes. Since the forecasting task will be done every 30 minutes, the whole system has enough time to execute the decomposition of the load sequence using VMD, pre-processing all sequences, and running the forecasting model on the inputs. A simple micro-controller with enough memory storage should be able to withstand the computational costs related to all operations of the proposed forecasting technique.

\section{Conclusion}\label{Sec:5}
This paper investigates the effectiveness of deep learning relying on VMD and mWDN to forecast short-term multi-horizon residential load consumption when no exogenous variables are available. To decompose load sequences into stationary sub-sequences, VMD is employed as the initial time-frequency analysis and wavelet transform as the subsequent decomposition within the mWDN model to extract various multi-resolution consumption patterns inherent in the decomposed sub-sequences. The proposed technique is compared with existing methods, including historical mean and other state-of-the-art deep learning models. In addition, the forecasting performance of the models is evaluated with and without VMD decomposition in order to validate its use for this application. Moreover, the impact of the decomposition level on the forecasting performances is also examined.

Five Moroccan households of different characteristics and at different times of the year were monitored for whole-house electricity consumption, using which the developed models are assessed. Results indicate that the proposed technique is the most effective of the considered methods. The fact that this technique outperforms its individual counterpart (VMD-only or mWDN-only) proves its effectiveness in capturing essential information describing the consumption behaviors of the premises' residents. A VMD decomposition level of $K=127$ and an mWDN decomposition level of $I=4$ provided the highest performance between all considered levels within the proposed technique and employed data. 

Future work will focus on continuing the data acquisition campaign in more Moroccan households and investigating very-short forecasting horizons.

\section*{Funding}
This work is partly funded by the USAID under the grant agreement number 2000007744 (PVBUILD project), by IRESEN for the MORESOLAR project, and the European Union’s Horizon 2020 research and innovation program AERIAL-CORE under grant agreement number 87147.

\section*{Declaration of Competing Interest}
The authors declare that they have no known competing financial interests or personal relationships that could have appeared to influence the work reported in this paper.

\bibliographystyle{elsarticle-num} 
\bibliography{cas-refs}

\end{document}